\documentclass[preprint,5p,number,sort&compress]{elsarticle}

\usepackage{amsmath}
\usepackage[T1]{fontenc}
\usepackage{url}
\usepackage{hyperref}
\usepackage{color}
\hypersetup{colorlinks=true}
\usepackage{bm}
\usepackage{amscd, mathrsfs}
\usepackage{amsthm}
\usepackage{cases}
\usepackage{here}
\usepackage{graphics}
\usepackage{ulem}
\usepackage{multirow}
\usepackage{subcaption}
\usepackage{graphicx}
\usepackage{amssymb}
\usepackage{tabularx}

\allowdisplaybreaks

\newcommand{\eqr}{\;\;=\hspace{-7.5mm}\raisebox{1.3ex}[0ex][1ex]{{\tiny{($\epsilon_0=1$)}}}}
\newcommand{\eqir}{\;\;=\hspace{-7.5mm}\raisebox{1.3ex}[0ex][1ex]{{\tiny{($4\pi\epsilon_0=1$)}}}}

\journal{Phys.~Lett.~B}
\begin{document}
\begin{frontmatter}
\title{Revisiting constraints on primordial magnetic fields from\\ spectral distortions of cosmic microwave background}
\author[1]{Fumio Uchida\corref{cor1}}
\ead{fuchida@post.kek.jp}
\cortext[cor1]{Corresponding author}
\affiliation[1]{organization={Theory Center, Institute of Particle and Nuclear Studies (IPNS), High Energy Accelerator Research Organization (KEK)}, postcode={305-0801}, city={Tsukuba, Ibaraki}, country={Japan}}

\author[2,3,4]{Kohei Kamada}
\ead{kohei.kamada@ucas.ac.cn}
\affiliation[2]{organization={School of Fundamental Physics and Mathematical Sciences, Hangzhou Institute for Advanced Study, University of Chinese Academy of Sciences (HIAS-UCAS)},  city={310024 Hangzhou}, country={China}}
\affiliation[3]{organization={International Centre for Theoretical Physics Asia-Pacific (ICTP-AP)},city={Hangzhou/Beijing}, country={China}}
\affiliation[4]{organization={Research Center for the Early Universe, The University of Tokyo},city={Bunkyo-ku, Tokyo 113-0033}, country={Japan}}
\author[5]{Hiroyuki Tashiro}
\ead{hiroyuki@ed.sojo-u.ac.jp}
\affiliation[5]{Center for Education and Innovation, Sojo University, Nishi-ku, Kumamoto, 860-0082 Japan}
\begin{abstract}
    The magneto-hydrodynamic decay of primordial magnetic fields can distort the black-body spectrum of the cosmic microwave background~(CMB) by draining magnetic energy into thermal plasmas and photons.
    The current limits on CMB distortion place constraints on small-scale primordial magnetic fields.
    The constraints crucially depend on the decay laws of primordial magnetic fields. Recent numerical simulations reveal that non-linear effects play a siginificant role in the magnetic field decay
    although these effects are neglected in previous works.
    In this paper, by adopting a reconnection-driven turbulent decay as a non-linear evolution model, 
    we demonstrate the potential impact of non-linear effects on CMB spectral distortions.
    The reconnection-driven turbulent decay model is an analytical description which provides the consistent results with numerical simulation.
    Our results rule out magnetic fields with shorter coherence lengths.
    While the result is independent of the spectral index of the magnetic energy spectrum, it is influenced by the magnetic helicity fraction.
\end{abstract}
\end{frontmatter}
\section{Introduction}
\noindent
The physics of plasma in the early universe plays an essential role in cosmology, particularly in the context of magneto-hydrodynamics (MHD), which governs the dynamics of primordial magnetic fields coupled with the plasma. 
Earlier studies often assumed adiabatic evolution, suggesting that primordial magnetic fields remain frozen in comoving units after their generation.
However, it has been recognized that the dissipation of magnetic fields must be taken into account, even in the limit of large electric conductivity due to their interaction with the plasma fluid, which possesses finite viscosity \cite{1998PhRvD..57.3264J}.

A proper description of the decay dynamics of primordial magnetic fields enables us to quantitatively relate magnetogenesis scenarios \cite{1988PhRvD..37.2743T, 1992ApJ...391L...1R, 1992PhRvD..46.5346G, PhysRevLett.51.1488, 1989ApJ...344L..49Q, PhysRevD.50.2421, 1991PhLB..265..258V} to the observed void magnetic fields \cite{2010Sci...328...73N,2010MNRAS.406L..70T,2011ApJ...727L...4D,2018ApJS..237...32A,2023A&A...670A.145A,2024MNRAS.527L..95D}.
Additionally, this description allows us to draw quantitative implications from phenomena associated with the evolution of primordial magnetic fields (see {\it e.g.}, \cite{2020JCAP...05..039S,2024arXiv240519693K,2018MNRAS.474L..52S,1998PhRvD..57.2186G,2016PhRvD..93h3520F,2021JCAP...04..034K,1969Natur.223..938G,2018PhRvD..98h3518S,2022PhRvD.105l3502R,2016A&A...594A..19P,2024JCAP...07..086P,1996ApJ...468...28K,2024PhRvD.109d3520S,2023PASJ...75S.154M,2024ApJ...972..117Z}). 
In this paper, we focus specifically on how the spectral distortions \cite{1970Ap&SS...7....3S,1980ARA&A..18..537S,1995A&A...303..323B} of the cosmic microwave background (CMB) can impose constraints on the properties of primordial magnetic fields \cite{Jedamzik:1999bm,2005MNRAS.356..778S,2014JCAP...01..009K,2015PhRvD..92l3004W}.

When primordial magnetic fields decay in the early universe, the energy released heats the CMB photons.
At sufficiently high temperatures, the Hubble expansion is slow enough that the distribution of CMB photons relaxes into a black-body spectrum as the heat only raises the temperature, making observational detection challenging. (See, {\it e.g.}, Refs.~\cite{2018MNRAS.474L..52S,2019MNRAS.490.4419S}.)
However, at lower temperatures, a black-body spectrum is no longer maintained against the energy injection because the scattering processes between photons and electrons, which enforce this spectrum, fall out of equilibrium \cite{1966ApJ...145..560W,1969Ap&SS...4..301Z,1970Ap&SS...7...20S,1982A&A...107...39D,1993PhRvD..48..485H,2012MNRAS.425.1129C,2012A&A...543A.136K,2013MNRAS.434..352C}.
Consequently, the energy injection from the decaying magnetic fields leaves observable traces as distortions in the black-body spectrum \cite{Jedamzik:1999bm,2005MNRAS.356..778S,2014JCAP...01..009K,2015PhRvD..92l3004W}.

A precise description of the magnetic field decay is then undoubtedly essential for evaluating the extent to which the CMB spectrum is distorted, since it determines the amount of energy injected.
References \cite{Jedamzik:1999bm,2005MNRAS.356..778S,2014JCAP...01..009K} take into account the damping of linear MHD modes \cite{1998PhRvD..57.3264J} and a certain class of non-linear mode, namely the non-linear Alfv\'{e}n mode \cite{1998PhRvD..58h3502S}.
On the other hand, highly non-linear regimes were poorly understood and often neglected in the literature except for Ref.~\cite{2015PhRvD..92l3004W}, which is based on a previous understanding of non-linear regimes \cite{2004PhRvD..70l3003B}.
After a series of study on fully non-linear decay laws \cite{2004PhRvD..70l3003B,2014ApJ...794L..26Z, 2015PhRvL.114g5001B,2001PhRvE..64e6405C, Brandenburg:2016odr, Brandenburg+17, 2017PhRvE..96e3105R,2017MNRAS.472.1628P}, a clear physical interpretation of the non-linear decay dynamics was recently proposed.

The non-linear MHD of our interest can be understood by the reconnection-driven turbulence and the Hosking integral conservation~\cite{2021PhRvX..11d1005H,2023NatCo..14.7523H,2023PhLB..84338002U,2024arXiv240506194U}.\footnote{For other potential explanations on the non-linear dynamics of non-helical magnetic fields, see Refs.~\cite{2013PhRvD..87h3007K,2021MNRAS.501.3074B,2024OJAp....7E..75D}.}
Such a non-linear physics is relevant, as is implied by a large magnetic Reynolds number, and substantially affects the quantitative analysis on CMB spectral distortions.
In this paper, we explore the fully non-linear regime, previously overlooked \cite{Jedamzik:1999bm,2005MNRAS.356..778S, 2014JCAP...01..009K} and discuss its relevance in updating the constraints on primordial magnetic fields from CMB spectral distortions.

The rest of the paper is organized as follows.
In Sec.~\ref{sec:ReconnectionDrivenTurbulence}, we review the properties and evolution of the decaying primordial magnetic fields.
In Sec.~\ref{sec:CMBDistortion}, we derive formulae to estimate the CMB $\mu$- and $y$-distortions generated {\it before} the recombination epoch.
Constraint on primodial magnetic fields is given in Sec.~\ref{sec:Observation}, and we summarize the results in Sec.~\ref{sec:Discussion}.
Throughout the paper, we adopt the natural Heaviside--Lorentz unit (\ref{qppx:emunits}), where $c=\hbar=k_{\rm B}=\epsilon_0=1$, $e^{2}=4\pi\alpha$ with $\alpha\simeq1/137$ being the fine-structure constant, $1\,{\rm G}=1.95\times 10^{-20}\,{\rm GeV}^2$, and $1\,{\rm Mpc}=1.56\times 10^{38}\,{\rm GeV}^{-1}$, and use the values of cosmological parameters in Planck 2018 results~\cite{2020A&A...641A...6P}.

\section{Magnetic field decay\label{sec:ReconnectionDrivenTurbulence}}
\noindent
In this section, we review Refs.~\cite{2023NatCo..14.7523H,2023PhLB..84338002U,2024arXiv240506194U} to derive formulae that describe the decay laws of primordial magnetic fields.

\subsection{Reconnection-driven turbulence}
\begin{figure}[t]
        \begin{minipage}[h]{1.0\hsize}
            \includegraphics[keepaspectratio, width=0.98\textwidth]{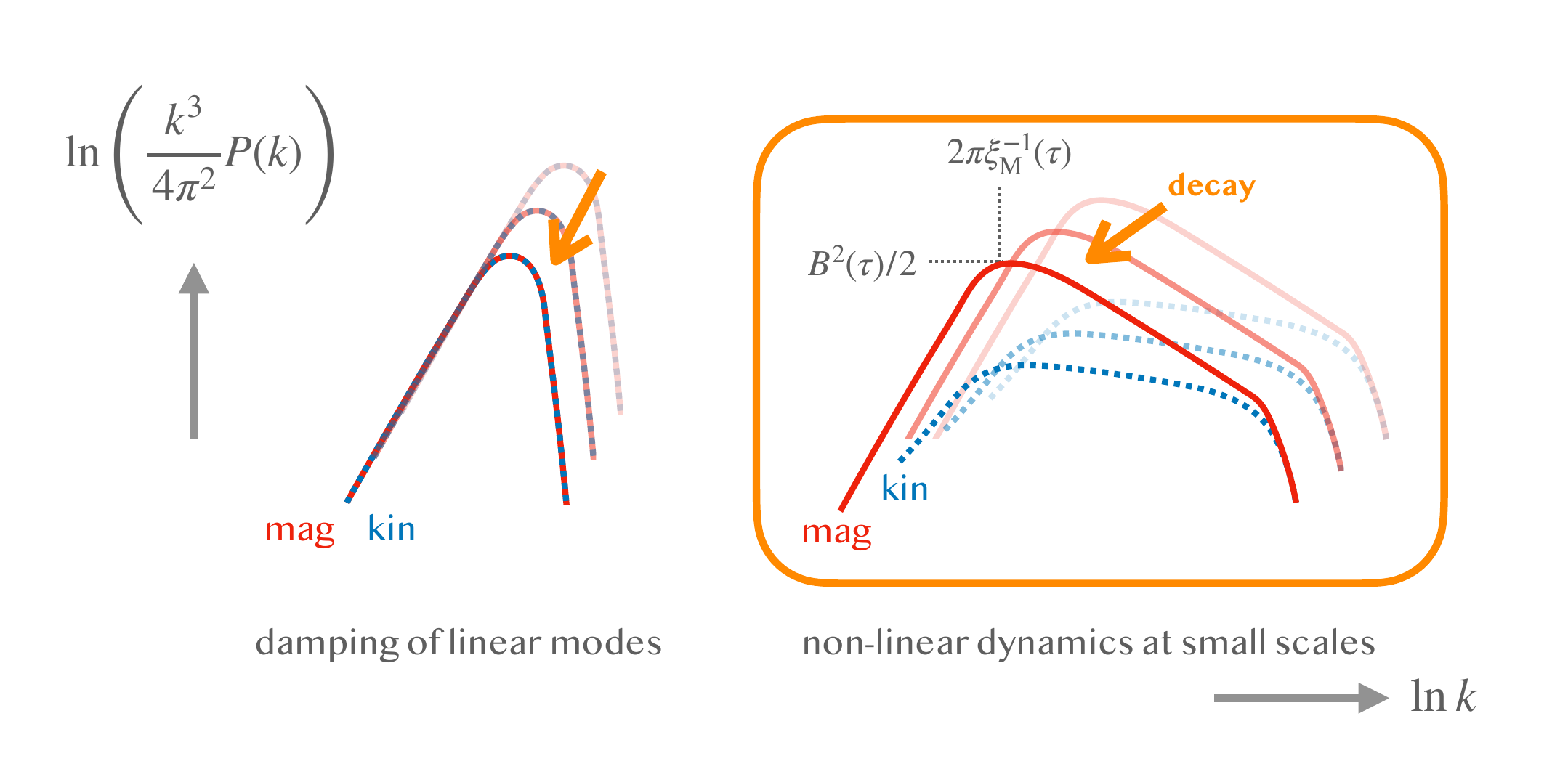}
        \end{minipage}
    \caption{\label{fig:MHD}Decay of magnetic (red solid) and kinetic (blue-dotted) energy spectra. The magnetic energy spectrum is $P(k,\tau)$ defined in Eq.~\eqref{eq:def_MES}, and the kinetic one is defined similarly for the velocity field of the plasma fluid, {\it i.e.}, $(\rho_{\rm fluid}+p_{\rm fluid})\langle{\boldsymbol v}({\boldsymbol k},\tau)\cdot{\boldsymbol v}({\boldsymbol k}',\tau)\rangle
        =(2\pi)^3\delta\left({\boldsymbol k}+{\boldsymbol k}'\right)P_{\rm kin}(k,\tau)$.} Left panel shows the damping of linear MHD modes considered in the previous literature \cite{2014JCAP...01..009K}. However, when the magnetic energy dominates, non-linear decaying processes invalidate the linear damping, which could be relevant at larger scales compared with the non-linear scale (Right panel; we refer figures in Ref.~\cite{2017PhRvD..96l3528B} for numerical results).
\end{figure}

\noindent
In the previous literature, primordial magnetic fields are supposed to damp at the same scales as the velocity field damps.
While the velocity field suffers the Silk damping at the Silk scale in the absence of the magnetic field \cite{1968ApJ...151..459S}, some ``overdamped'' modes that survive inside the Silk scale are found in the linear MHD \cite{1998PhRvD..57.3264J}.
Then, the damping scale of these overdamped modes is identified as the damping scale of primordial magnetic fields in the literature \cite{2014JCAP...01..009K}.
A schematic picture of this description is given in the left panel of Fig.~\ref{fig:MHD},
where the power spectra of the magnetic energy and the kinetic energy are exponentially damped at the same scale.

However, non-linear effects can modify the decay scenario of primordial magnetic fields.
Once non-linear effects become important, the energy of the primordial magnetic fields transfers to different scales due to the non-linear coupling in the MHD equations.\footnote{Note that, even if the overdamped modes determine the dissipation scale, non-linear dynamics can make the magnetic coherence length {\it larger} than the dissipation scale.}
Therefore, primordial magnetic fields may survive even {\it inside} the above-mentioned damping scales in magnetically-dominated regimes (Right panel in Fig.~\ref{fig:MHD}).

The primordial magnetic field of our interest has typical magnetic strength $B\gtrsim 1\,{\rm nG}$ and its coherence length $\xi_{\rm M}\gtrsim 10^{-9}\,{\rm Mpc}$. 
Such a strong magnetic field implies a large magnetic Reynolds number in the mostly radiation-dominated universe before the recombination
(\ref{appx:nonlinearity}),
\begin{align}
    {\rm Re}_{\rm M}\gg1,
    \label{eq:large_ReM}
\end{align}
and therefore we expect a turbulent decay of primordial magnetic fields \cite{2023NatCo..14.7523H,2024arXiv240506194U}.
In such a turbulent regime, the timescale of magnetic field decay is determined by the relevant process that transfers the magnetic energy into the velocity field and heat.
It was recently proposed that magnetic reconnection \cite{1957JGR....62..509P,1958IAUS....6..123S,1984PhFl...27..137P} drives such energy transfer due to magnetic dissipation inside a thin two-dimensional structure called current sheets \cite{2021PhRvX..11d1005H}.

In the next subsection, we provide the scaling relation for the evolution of the primordial magnetic fields resulting from the reconnection-driven turbulence, focusing on the typical magnetic field strength and coherence length.

\subsection{scaling solution}
\noindent
Throughout this paper, we assume that primordial magnetic fields ${\boldsymbol B}$ are statistically homogeneous and isotropic Gaussian random fields. 
The energy spectrum, $P(k)$, of such fields is given in
\begin{align}
    \langle{\boldsymbol B}({\boldsymbol k},\tau)\cdot{\boldsymbol B}({\boldsymbol k}',\tau)\rangle
        =(2\pi)^3\delta\left({\boldsymbol k}+{\boldsymbol k}'\right)P(k,\tau).
    \label{eq:def_MES}
\end{align}
where ${\boldsymbol B}({\boldsymbol k},\tau)$ is the Fourier component of comoving primordial magnetic fields and the angle bracket represents the ensemble average.
Instead of specifying $P(k,\tau)$, we parameterize the magnetic fields by the typical magnetic strength~$B(\tau)$ and the coherence length~$\xi_{\rm M}(\tau)$ to express the decay of the primordial magnetic fields. 
In terms of the power spectrum~$P(k)$, $B$ and $\xi_{\rm M}$ are defined as
\begin{align}
    B^2
        &:=2\langle \rho_B\rangle=\int\dfrac{d^3k}{(2\pi)^3}P(k), \label{Bdef} \\
    \xi_{\rm M}
        &:=\dfrac{1}{B^2}\int\dfrac{d^3k}{(2\pi)^3}\dfrac{2\pi}{k}P(k)\label{xidef},
\end{align}
where $\rho_B$ is the comoving energy density for primordial magnetic field, and we have omitted the argument $\tau$.

To examine the time evolution of the magnetic fields, we need to determine their typical time scale in 
terms of these parameters, $B$ and $\xi_{\rm M}$.
In reconnection-driven turbulence, which is the case of our interest, we identify it as the reconnection timescale.
Among the multiple regimes of magnetic reconnection processes \cite{2011PhPl...18k1207J}, the so-called fast magnetic reconnection is the relevant one because the strong magnetic field of our interest implies a large Lundquist number $S\gg S_{\rm c}$, with $S_{\rm c}\sim 10^4$ being the numerically established critical value \cite{2011PhPl...18k1207J,1986PhFl...29.1520B,2005PhRvL..95w5003L,2007PhPl...14j0703L}, implies (\ref{appx:nonlinearity}).
The timescale of the fast magnetic reconnection is given as \cite{2023NatCo..14.7523H,2024arXiv240506194U}
\begin{align}
    \tau_{\rm fast}
        :=S_{\rm c}^{\frac{1}{2}}(\rho_{\rm fluid}+p_{\rm fluid})^{\frac{1}{2}}\sigma^{\frac{1}{2}}\eta^{\frac{1}{2}}B^{-1}\xi_{\rm M},
        \label{eq:timescale_fast}
\end{align}
where $\rho_{\rm fluid}$ and $p_{\rm fluid}$ are the comoving energy density and pressure of the plasma fluid coupled with the magnetic field, respectively, $\sigma$ is the comoving electric conductivity, and $\eta$ is the comoving shear viscosity.
This timescale imposes a condition for primordial magnetic fields to survive~\cite{2024arXiv240506194U}
\begin{align}
    \tau
        \leq\tau_{\rm fast}\left(\tau,B,\xi_{\rm M}\right),
        \label{eq:timescalecondition}
\end{align}
where the equality holds when the magnetic field is in the process of reconnection-driven decay.
If the equality does not hold, the magnetic field is frozen.
During the process of reconnection-driven decay,
from Eqs.~\eqref{eq:timescale_fast} and \eqref{eq:timescalecondition} it follows 
\begin{align}
    B\xi_{\rm M}^{-1}
        =S_{\rm c}^{\frac{1}{2}}(\rho_{\rm fluid}+p_{\rm fluid})^{\frac{1}{2}}\sigma^{\frac{1}{2}}\eta^{\frac{1}{2}}\tau^{-1}.
        \label{eq:cond_decaying}
\end{align}
It determines magnetic field strength in the decaying process as a function of magnetic coherence length,
\begin{align}
    \dfrac{B}{10^{-7}\,{\rm G}}
        &=\left\{\begin{matrix}\vspace{3mm}
        4\cdot\dfrac{\xi_{\rm M}}{10^{-3}\,{\rm Mpc}}\cdot\dfrac{z}{10^3}&\,\text{(radiation-dominated),}\\
        8\cdot\dfrac{\xi_{\rm M}}{10^{-3}\,{\rm Mpc}}\hspace{8mm}\text{}&\text{(at the recombination),}
        \end{matrix}\right.\label{eq:BXi_relation}
\end{align}
where we properly take into account the matter-dominated epoch, $z<z_{\rm eq}=3400$ \cite{2020A&A...641A...6P}, before the recombination.

Another quantity that characterizes the primordial magnetic fields is the ensemble average of magnetic helicity density, 
$\langle {\bm A} \cdot {\bm B} \rangle$, where ${\bm A}$ is the vector potential. 
Note that the ensemble average is practically identified as a volume average beyond all the characteristic length scales.
This average of magnetic helicity density is then gauge invariant if no magnetic fluxes penetrate the spatial boundary.
Also, if the coherence length of the magnetic field is larger than the magnetic dissipation scale, $\sigma^{-1/2}\tau^{1/2}$, which holds in non-linear regimes \cite{2024arXiv240506194U}, it behaves as a good conserved quantity \cite{2021PhRvX..11d1005H}.
We introduce a parameter called the helicity fraction, $\epsilon$, such that
\begin{align}
    \langle{\boldsymbol{A}}\cdot{\boldsymbol{B}}\rangle
        =\dfrac{\epsilon}{2\pi} B^2\xi_{\rm M},
\end{align}
to quantify the parity violation in the magnetic field configuration.
The Schwartz inequality bounds $\vert \epsilon\vert\leq1$ \cite{1978mfge.book.....M}, meaning the magnetic field is maximally helical when $\epsilon=\pm1$.
The opposite limit, $\vert\epsilon\vert\ll1$, corresponds to a non-helical magnetic field. 

The evolution of the primordial magnetic fields crucially depends on whether the magnetic field is helical or not.
When the primordial magnetic field is maximally helical, magnetic helicity conservation in the large conductivity limit imposes a constraint \cite{1975JFM....68..769F}
\begin{align}
    B^2\xi_{\rm M}
        ={\rm const.}\hspace{10mm}
        \text{(maximally helical)}
        \label{eq:max}
\end{align}
which can be solved with Eq.~\eqref{eq:cond_decaying} to imply \cite{2024arXiv240506194U}
\begin{align}
    B
        &=B_{\rm ini}^{\frac{2}{3}}\xi_{\rm M,ini}^{\frac{1}{3}}S_{\rm c}^{\frac{1}{6}}(\rho_{\rm fluid}+p_{\rm fluid})^{\frac{1}{6}}\sigma^{\frac{1}{6}}\eta^{\frac{1}{6}}\tau^{-\frac{1}{3}}\notag\\
        &\hspace{39mm}\text{(maximally helical)} \label{eq:max_b}\\
        &=
        B_{\rm ini}\left(\dfrac{\tau}{\tau_{\rm ini}}\right)^{-\frac{1}{3}}\hspace{10.5mm}(\tau>\tau_\mathrm{ini}),\label{eq:max_b_power}\\
    \xi_\mathrm{M} &=B_{\rm ini}^{\frac{2}{3}}\xi_{\rm M,ini}^{\frac{1}{3}}S_{\rm c}^{-\frac{1}{3}}(\rho_{\rm fluid}+p_{\rm fluid})^{-\frac{1}{3}}\sigma^{-\frac{1}{3}}\eta^{-\frac{1}{3}}\tau^{\frac{2}{3}}\notag\\
        &\hspace{39mm}\text{(maximally helical)} \label{eq:max_xi}\\   
        &=
        \xi_{\rm M,ini}\left(\dfrac{\tau}{\tau_{\rm ini}}\right)^{\frac{2}{3}}\hspace{10.5mm}(\tau>\tau_\mathrm{ini}). 
\end{align}
In the second line, we assume that the magnetic field enters the decaying regime at $\tau = \tau_\mathrm{ini}$. 
In other words, $\tau_\mathrm{ini}$ is determined such that Eqs.~\eqref{eq:max_b} and \eqref{eq:max_xi} are satisfied as $B (\tau_\mathrm{ini})=B_\mathrm{ini}$ and $\xi_\mathrm{M}(\tau_\mathrm{ini})=\xi_\mathrm{M,ini}$.

On the contrary, when the primordial magnetic field is non-helical, the global magnetic helicity conservation $\epsilon B^2\xi_{\rm M}={\rm const.}\simeq 0$ does not constrain the magnetic field.
Previously, it was commonly assumed that large-scale modes of the magnetic energy spectrum are conserved \cite{2004PhRvD..70l3003B,2014JCAP...01..009K}, and therefore the decay law of the non-helical primordial magnetic fields was thought to be determined by the magnetic spectral index.
See the left panel of Fig.~\ref{fig:MHD}.
However, the assumption can be violated when small-scale modes non-linearly affect large-scale modes.
Indeed, recent numerical studies have observed the inverse transfers, clear contradictions to the previous expectation \cite{2014ApJ...794L..26Z,2015PhRvL.114g5001B,2001PhRvE..64e6405C,2017PhRvD..96l3528B,2017PhRvE..96e3105R,2017MNRAS.472.1628P}.

Subsequently, a quantity called the Hosking integral, which quantifies local fluctuations of magnetic helicity, was proposed \cite{2021PhRvX..11d1005H}. 
It is considered conserved due to the conservation of magnetic helicity within finite subvolumes.
Its conservation has already been confirmed by follow-up numerical studies \cite{2022JPlPh..88f9002Z,2023Atmos..14..932B,2023JPlPh..89f9006B,2024arXiv240611798B}.
Based on dimensional analysis, the Hosking integral scales as $B^4\xi_{\rm M}^5$, and therefore we obtain a conservation law (\ref{appx:HI}), 
\begin{align}
    B^4\xi_{\rm M}^5
        ={\rm const.}\hspace{10mm}
        \text{(non-helical)}. 
        \label{eq:non-}
\end{align}
With Eq.~\eqref{eq:cond_decaying}, it is implied \cite{2024arXiv240506194U}
\begin{align}
    B
        &=B_{\rm ini}^{\frac{4}{9}}\xi_{\rm M,ini}^{\frac{5}{9}}S_{\rm c}^{\frac{5}{18}}(\rho_{\rm fluid}+p_{\rm fluid})^{\frac{5}{18}}\sigma^{\frac{5}{18}}\eta^{\frac{5}{18}}\tau^{-\frac{5}{9}}\notag\\
        &\hspace{53.5mm}\text{(non-helical)}
        \label{eq:non_b}\\
        &=B_{\rm ini}\left(\dfrac{\tau}{\tau_{\rm ini}}\right)^{-\frac{5}{9}}\hspace{15mm}(\tau>\tau_\mathrm{ini}),\label{eq:non_b_power}\\
    \xi_\mathrm{M}
        &=B_{\rm ini}^{\frac{4}{9}}\xi_{\rm M,ini}^{\frac{5}{9}}S_{\rm c}^{-\frac{2}{9}}(\rho_{\rm fluid}+p_{\rm fluid})^{-\frac{2}{9}}\sigma^{-\frac{2}{9}}\eta^{-\frac{2}{9}}\tau^{\frac{4}{9}} \notag\\
        &\hspace{53.5mm}\text{(non-helical)}
        \label{eq:non_xi} \\
        &=\xi_{\rm M,ini}\left(\dfrac{\tau}{\tau_{\rm ini}}\right)^{\frac{4}{9}}\hspace{10.5mm}(\tau>\tau_\mathrm{ini}),         
\end{align}
In the second line, we assume that the magnetic field enters the decaying regime at $\tau=\tau_{\rm ini}$.
In other words, once more, $\tau_\mathrm{ini}$ is determined such that Eqs.~\eqref{eq:non_b} and \eqref{eq:non_xi} are satisfied as $B(\tau_\mathrm{ini})=B_\mathrm{ini}$ and $\xi_\mathrm{M}(\tau_\mathrm{ini})=\xi_\mathrm{M,ini}$.
By solving Eq.~\eqref{eq:cond_decaying} and either Eq.~\eqref{eq:max} or \eqref{eq:non-}, we identify the magnetic field strength and coherence length at a given time for given initial conditions.

One might expect that magnetic coherence lengths obtained in this way should be larger than the damping scale of the overdamped MHD modes \cite{1998PhRvD..57.3264J,1998PhRvD..58h3502S}.
However, this is not the case; smaller-scale magnetic fields can survive in the reconnection-driven turbulence, as already explained in the beginning of this section.
This is because multiple modes at different scales are involved in the non-linear reconnection turbulence,\footnote{This can be understood by a simple argument based on two velocities that characterize the turbulence. A velocity $v_{\rm out}$ balances with the magnetic field in the presence of photon drag, in the sense $v_{\rm A}^2/\xi_{\rm M}\sim v_{\rm out}/Rl_\gamma$, where $v_{\rm A}:=B/\sqrt{\rho_{\rm fluid}+p_{\rm fluid}}$ is the Alfv\'{e}n velocity, $R:=3\rho_{\rm b}/4\rho_\gamma$ is the baryon-to-photon enthalpy ratio, and $l_{\gamma}$ is the photon mean free path.
Another velocity $v_{\rm in}$ controls the timescale, in the sense $\tau_{\rm fast}\sim\xi_{\rm M}/v_{\rm in}$.
If $v_{\rm in}\sim v_{\rm out}$, the condition \eqref{eq:timescalecondition} implies damping roughly inside the scale obtained in Refs.~\cite{1998PhRvD..57.3264J,1998PhRvD..58h3502S}. 
However, with a hierarchy $v_{\rm in}\ll v_{\rm out}$, much smaller $\xi_{\rm M}$ can be consistent with the condition \eqref{eq:timescalecondition}.
We refer readers to Refs.~\cite{2021PhRvX..11d1005H,2023NatCo..14.7523H,2023PhLB..84338002U,2024arXiv240506194U} for more explanation on the reconnection-driven turbulence.} whereas the aforementioned damping scales are obtained for a single planar-wave ansatz, which does not work at all in the present situation.

\section{CMB distortion by decaying magnetic field\label{sec:CMBDistortion}}
\noindent
In this section, we apply the scaling evolution of the primordial magnetic fields shown in the previous section to estimate the CMB spectral distortions due to the energy injection from the decaying primordial magnetic fields before the recombination.\footnote{In the formulae developed in Refs.~\cite{2023PhLB..84338002U,2024arXiv240506194U}, primordial magnetic fields are approximated to be frozen after recombination and thus do not contribute to CMB spectral distortions. However, as discussed in Ref.~\cite{2005MNRAS.356..778S}, magnetic field may continue to undergo turbulent decay. Addressing this would require incorporating the Hosking integral and reconnection time-scale arguments also in this regime, which is beyond the scope of the present study. Therefore we do not investigate the CMB distortion occuring after recombination here.}

\subsection{the energy injection rate}
\noindent
As discussed in the previous section, the primordial magnetic fields undergo non-linear decay.
The coherence length of the primordial magnetic fields, $\xi_\mathrm{M}$, with a typical field strength peaked at this scale, $B$, evolves according to Eq.~\eqref{eq:max_xi} for maximally helical magnetic fields and Eq.~\eqref{eq:non_xi} for non-helical magnetic fields. 
This behavior indicates an apparent energy transfer to larger scales. 
However, at the same time, some fraction of magnetic field energy is transferred to smaller scales, where it is subsequently damped into the plasma on the scale of the magnetic field dissipation.

The comoving energy injection rate, $d\varepsilon/d\tau$, from the primordial magnetic fields into plasma can be estimated from the decay rate of comoving magnetic energy as
\begin{align}
    \dfrac{d\varepsilon}{d\tau}
        &=-\dfrac{d \langle \rho_B \rangle}{d\tau}
        =\dfrac{\alpha+1}{2}\dfrac{B^2_{\rm ini}}{\tau_{\rm ini}}\left(\frac{\tau}{\tau_\mathrm{ini}}\right)^{-\alpha-2},\label{eq:dQdz_rad}
\end{align}
where we define
\begin{align}
    \alpha
        =\left\{ \begin{array}{l}
        -\dfrac{1}{3}\hspace{10mm}\text{(maximally helical)}\\[9pt]
        \dfrac{1}{9}\hspace{23mm}\text{(non-helical)}
        \end{array}\right..\label{eq:alpha_def}
\end{align}
Here we have used Eqs.~\eqref{eq:max_b_power} and \eqref{eq:non_b_power}.

\subsection{the CMB distortions}
\noindent
In terms of CMB distortions, the early universe before the recombination can be decomposed into three different epochs \cite{1993PhRvD..48..485H,2013MNRAS.434..352C}. In the earliest epoch, $z>z_{\rm th}=2\times 10^6$, is the so-called ``$T$-era'', where the thermalization process of photons due to the double Compton scattering is efficient.
In this epoch, even though the energy is injected into the CMB, the blackbody spectrum can be maintained \cite{1970Ap&SS...7...20S,1993PhRvD..48..485H}.

The second epoch, $z_{\rm th}>z>z_{\rm BE}=5\times 10^4$, is the so-called ``$\mu$-era'', where the Compton scattering is efficient.
Once the energy is injected, the efficient Compton scattering leads CMB photons with injected energy to the Compton equilibrium,
which is described by the Bose--Einstein spectrum with non-zero chemical potential \cite{1970Ap&SS...7...20S,1982A&A...107...39D,1991A&A...246...49B,1993PhRvD..48..485H,2013MNRAS.434..352C}. 
This spectral distortion is called $\mu$-distortion.

The third epoch, $z_{\rm BE}<z<z_{\rm rec}$, where $z_{\rm rec}\simeq1100$ \cite{2020A&A...641A...6P} is the redshift at the recombination epoch, is the $y$-era.
Since Compton scattering is inefficient in this epoch, the CMB photons do not reach the equilibrium spectrum with the injected energy.
Instead, CMB photons are suffered by only a few Compton scatterings with the injected energy, and the resultant spectrum is described as $y$-distortions with the Compton $y$ parameter \cite{1966ApJ...145..560W,1969Ap&SS...4..301Z,1991A&A...246...49B,1993PhRvD..48..485H,2013MNRAS.434..352C}.

First, we evaluate the $\mu$-distortion. 
We use conformal time as the time coordinate instead of redshift to simplify the expressions.
Here we define $\tau_{\rm th}:=\tau(z_{\rm th})$ and $\tau_{\rm BE}:=\tau(z_{\rm BE})$, where $\tau(z)=5\times10^{19}\,{\rm s}\,z^{-1}$ (so that the conformal time element $d\tau$ coincides with the physical time element today) in the radiation-dominated epoch.
The chemical potential~$\mu$ in $\mu$-distortion is evaluated by \cite{2012MNRAS.425.1129C,2012A&A...543A.136K,2013MNRAS.434..352C,2014JCAP...01..009K}
\begin{align}
    \mu
        =\dfrac{1.4}{3}\int_{0}^{\tau_{\rm BE}}d\tau\dfrac{\frac{d\varepsilon}{d\tau}}{\rho_\gamma}e^{-\left(\frac{\tau}{\tau_{\rm th}}\right)^{-\frac{5}{2}}}.\label{eq:mu_general_tau}
\end{align}
Here $\rho_\gamma$ is the comoving energy density of radiation.

By substituting Eq.~\eqref{eq:dQdz_rad} into Eq.~\eqref{eq:mu_general_tau}, $\mu$ can be rewritten as 
\begin{align}
    \mu
        &=\dfrac{1.4}{3}\dfrac{\alpha+1}{5}G_\alpha(\tau_{\rm ini})\left(\dfrac{\tau_{\rm ini}}{\tau_{\rm th}}\right)^{\alpha+1}\dfrac{B^2_{\rm ini}}{\rho_\gamma},
\end{align}
where
\begin{align}
    G_\alpha(\tau)
        &:=\Gamma\left(\frac{2}{5}(\alpha+1),\left(\frac{\tau_{\rm th}}{\tau_{\rm BE}}\right)^{\frac{5}{2}}\right)\notag\\
        &\hspace{16mm}-\Gamma\left(\frac{2}{5}(\alpha+1),\left(\frac{\tau_{\rm th}}{\tau}\right)^{\frac{5}{2}}\right),
\end{align}
and $\Gamma$ is the incomplete gamma function.
Here $\alpha$ is a parameter defined in Eq.~\eqref{eq:alpha_def} that characterizes the redshift evolution of typical magnetic amplitude.
We introduce $c^\mu_\alpha(\tau):=G_\alpha(\tau)/G_\alpha(0)$, in terms of which
\begin{align}
    \mu
        &=7\times 10^{-5}\,\left(\dfrac{B_{\rm ini}}{10^{-7}\,{\rm G}}\right)^{\frac{4}{3}}\left(\dfrac{\xi_{\rm M,ini}}{10^{-8}\,{\rm Mpc}}\right)^{\frac{2}{3}}\notag\\
        &\quad\times c^{\mu}_{-1/3}(\tau_{\rm ini}),
        \label{eq:max_mu}
\end{align}
for maximally helical fields and
\begin{align}
    \mu
        &=2\times10^{-5}\,\left(\dfrac{B_{\rm ini}}{10^{-7}\,{\rm G}}\right)^{\frac{8}{9}}\left(\dfrac{\xi_{\rm M,ini}}{10^{-8}\,{\rm Mpc}}\right)^{\frac{10}{9}}\notag\\
        &\quad\times c^{\mu}_{1/9}(\tau_{\rm ini}),
        \label{eq:non_mu}
\end{align}
for non-helical fields, where we have used the relation between $\tau_{\rm ini}$ and $\xi_{\rm M,ini}$ in the first line of Eq.~\eqref{eq:BXi_relation}.

After the elastic Compton scattering becomes inefficient, the energy injection into photons contributes as the $y$-distortion \cite{2012MNRAS.425.1129C,2012A&A...543A.136K,2013MNRAS.434..352C,2014JCAP...01..009K}, 
\begin{align}
    y
        &=\dfrac{1}{12}\int_{\tau_{\rm BE}}^{\tau_{\rm rec}}d\tau\,\dfrac{\frac{d\varepsilon}{d\tau}}{\rho_\gamma}.
        \label{eq:y_general}
\end{align}
We can evaluate Eq.~\eqref{eq:y_general} by substituting the first lines of Eqs.~\eqref{eq:max_b} and \eqref{eq:non_b},
\begin{align}
    y
        &=2\times10^{-5}\,\left(\dfrac{B_{\rm ini}}{10^{-7}\,{\rm G}}\right)^{\frac{4}{3}}\left(\dfrac{\xi_{\rm M,ini}}{10^{-6}\,{\rm Mpc}}\right)^{\frac{2}{3}}\notag\\
        &\quad\times c^y_{2/3}(\max\{\tau_{\rm ini}, \tau_{\rm BE}\}),
        \label{eq:max_y}
\end{align}
for maximally helical magnetic fields and
\begin{align}
    y
        &=1\times10^{-5}\,\left(\dfrac{B_{\rm ini}}{10^{-7}\,{\rm G}}\right)^{\frac{8}{9}}\left(\dfrac{\xi_{\rm M,ini}}{10^{-6}\,{\rm Mpc}}\right)^{\frac{10}{9}}\notag\\
        &\quad\times c^y_{10/9}(\max\{\tau_{\rm ini}, \tau_{\rm BE}\}),
        \label{eq:non_y}
\end{align}
for non-helical magnetic fields, where we have introduced $c^y_{\alpha+1}(\tau):=(\tau^{-(\alpha+1)}_{\rm rec}-\tau^{-(\alpha+1)})/(\tau^{-(\alpha+1)}_{\rm rec}-\tau^{-(\alpha+1)}_{\rm BE})$, so that $c^y_{\alpha+1}(\tau_{\rm BE})=1$.
Note that we ignore the additional $y$-distortion that may be generated after recombination.

Note also that we assume $\tau_{\rm ini}<\tau_{\rm BE}$ for $\mu$-distortions and $\tau_{\rm ini}<\tau_{\rm rec}$ for $y$-distortions in the above derivation.
If primordial magnetic fields are generated at scales large enough to remain frozen until $\tau_{\rm BE}$ or $\tau_{\rm rec}$, or if they are generated in the late universe, our formulae do not apply.

\section{Observational constraints\label{sec:Observation}}
\begin{figure}[ht]
        \begin{minipage}[h]{1.0\hsize}
            \includegraphics[keepaspectratio, width=0.96\textwidth]{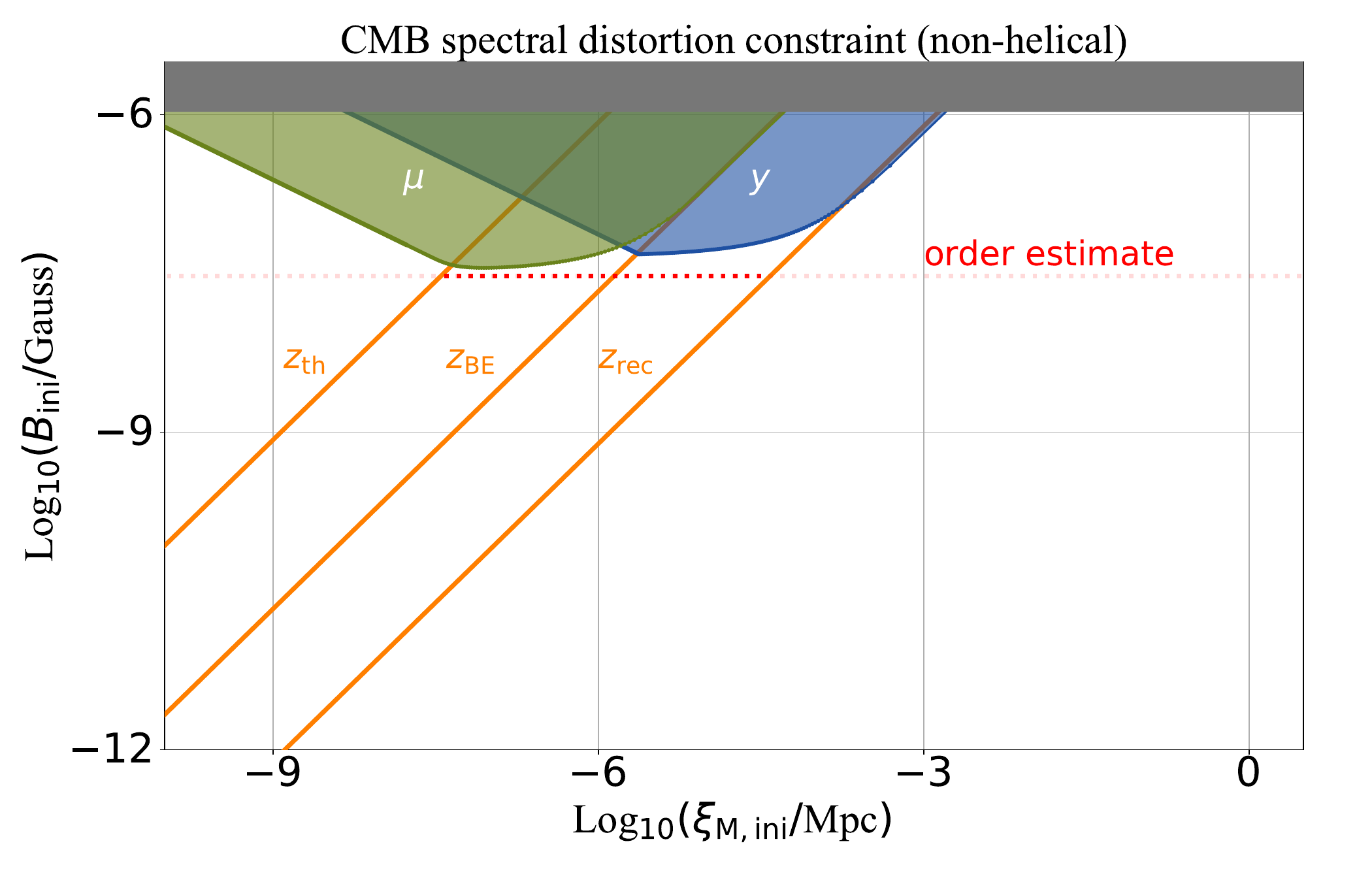}
        \end{minipage}
        \begin{minipage}[h]{1.0\hsize}
            \includegraphics[keepaspectratio, width=0.96\textwidth]{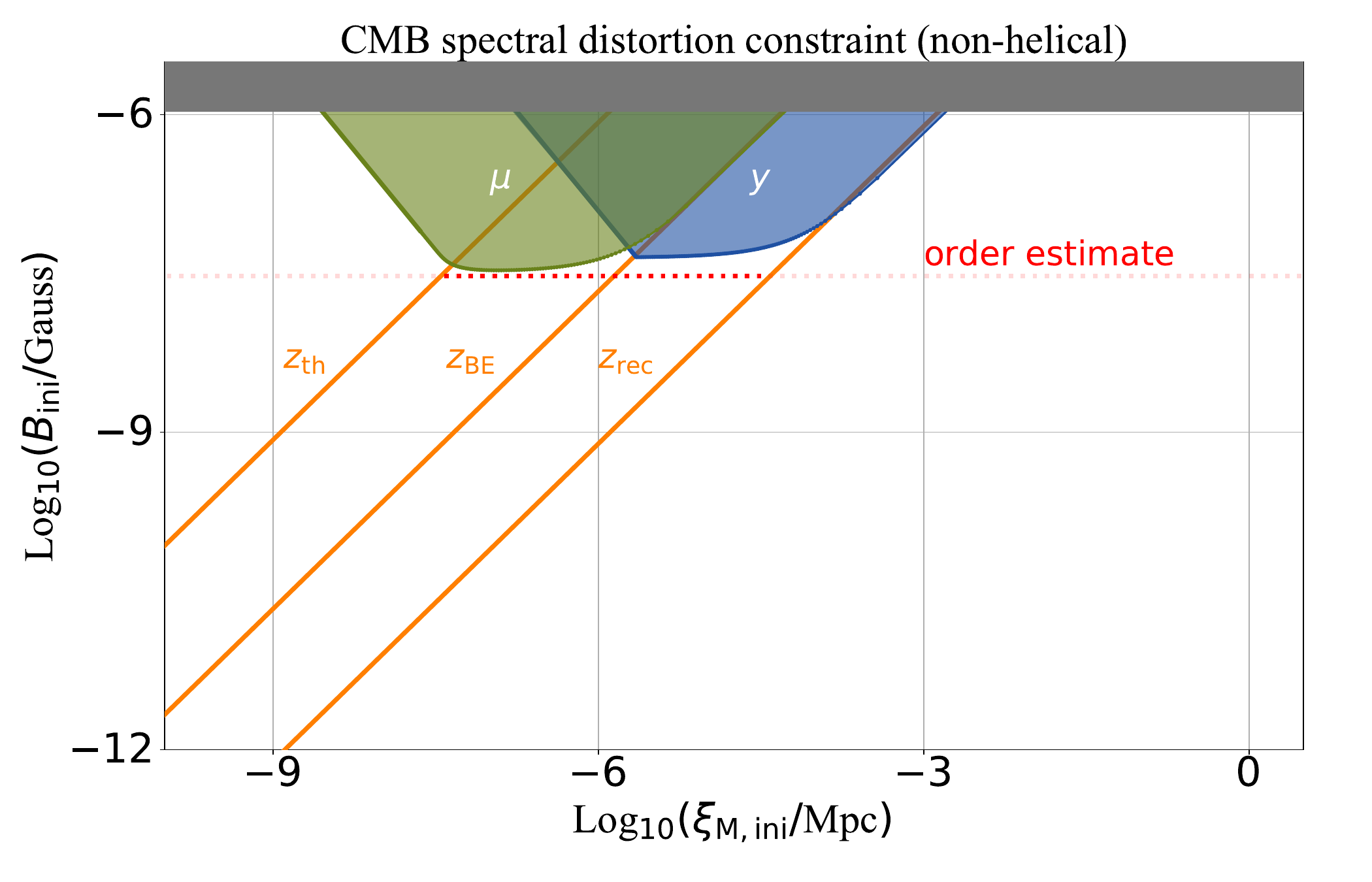}
        \end{minipage}
    \vspace{8mm}
    \caption{\label{fig:constraint}Constraints on comoving strength $B_{\mathrm{ini}}$ and coherence length $\xi_{\rm M, ini}$ of maximally helical ({\it Top}) and non-helical ({\it Bottom}) primordial magnetic fields at the time when they enter the decaying regime are shown. Green-shaded regions are excluded by the CMB constraint on $\mu$-distortion, and blue-shaded ones are by $y$-distortion. Inside gray-shaded regions, magnetic energy surpasses the total radiation energy. For reference, we show the order estimate \eqref{eq:oestimate} with the red-dotted lines and the $B-\xi_\mathrm{M}$ relation for decaying magnetic field \eqref{eq:cond_decaying} at $z_{\rm th}$, $z_{\rm BE}$, and $z_{\rm rec}$ with the apricot solid lines.}
\end{figure}

\noindent
In this section, we constrain primordial magnetic fields from the current observational bound on the $\mu$- and $y$-distortions.
The observational constraints are\footnote{We adopt the recently updated constraint on $\mu$ \cite{2022PhRvD.106f3527B}, which incorporates a refined analysis of the Galactic foreground modeling. We thank the anonymous reviewer for bringing it to our attention.} \cite{1996ApJ...473..576F,2022PhRvD.106f3527B}
\begin{align}
    \vert\mu\vert<4.7\times10^{-5},\quad
    \vert y\vert<1.5\times10^{-5}.
    \label{eq:obs}
\end{align}
By requiring that $B_{\rm ini}$ and $\xi_{\rm M,ini}$ satisfy Eqs.~\eqref{eq:cond_decaying}, \eqref{eq:max_mu} or \eqref{eq:non_mu}, \eqref{eq:max_y} or \eqref{eq:non_mu}, and \eqref{eq:obs} at every $z_{\rm ini}$, we exclude the green and blue shaded regions in Fig.~\ref{fig:constraint}.
In what follows, we provide a simplified approximation of the excluded regions.

If we approximate that a substantial portion of magnetic energy is injected into CMB photons, we obtain
\begin{align}
    \mu
        &\sim\dfrac{B^2(\max\{\tau_{\rm ini}, \tau_{\rm th}\})/2}{\rho_\gamma},
    \label{eq:mu_oestimate}\\
    y
        &\sim\dfrac{B^2(\max\{\tau_{\rm ini},\tau_{\rm BE}\})/2}{\rho_\gamma}.
    \label{eq:y_oestimate}
\end{align}
With these order estimates, Eqs.~\eqref{eq:mu_oestimate} and \eqref{eq:y_oestimate}, the observational constraints $\mu,\,y\lesssim10^{-5}$ correspond to
\begin{align}
    B_{\mathrm{ini}}\lesssim3\times10^{-8}\,{\rm G},
    \label{eq:oestimate}
\end{align}
which excludes magnetic fields at $z_{\rm th}>z> z_{\rm rec}$ above the red-dotted lines in Fig.~\ref{fig:constraint}.
The conditions imposed on the decay timescales, Eq.~\eqref{eq:cond_decaying}, which determines the relation between the magnetic field strength and coherence length during the turbulent decay,
at $z=z_{\rm th}$ and $z=z_{\rm BE}$ are shown as well by the apricot solid lines.
Since decaying primordial magnetic fields at $z=z_{\rm th},\,z_{\rm BE}$ and $z_{\rm rec}$ are on the $z_{\rm th}$, $z_{\rm BE}$, and $z_{\rm rec}$ lines, respectively, the red-dotted lines put an upper bound on $B_{\rm ini}$ when $\xi_{\rm M,ini}$ lies between the $z_{\rm th}$ and $z_{\rm rec}$ lines.
As expected, the lower boundaries of the excluded regions in Fig.~\ref{fig:constraint} roughly align with the red-dotted lines.

If we assume $z_{\rm ini}>z_{\rm th}$, {\it i.e.}, the primordial magnetic field enters the decaying regime before the $\mu$-era, we use Eqs.~\eqref{eq:max_mu} and \eqref{eq:non_mu} to obtain
\begin{align}
    \left(\dfrac{B_{\rm ini}}{10^{-7}\,{\rm G}}\right)^{2}\left(\dfrac{\xi_{\rm M,ini}}{10^{-8}\,{\rm Mpc}}\right)<6\times10^{-1},
    \label{eq:mu_before_max}
\end{align}
for maximally helical fields and
\begin{align}
    \left(\dfrac{B_{\rm ini}}{10^{-7}\,{\rm G}}\right)^{4}\left(\dfrac{\xi_{\rm M,ini}}{10^{-8}\,{\rm Mpc}}\right)^5<2\times10^1,
    \label{eq:mu_before_non}
\end{align}
for non-helical fields, from the $\mu$ constraint.
These conditions define the left-side edges of the green shaded regions in Fig.~\ref{fig:constraint}.

In a parallel manner, for the $y$ constraint, we assume $z_{\rm ini}>z_{\rm BE}$ and use Eqs.~\eqref{eq:max_y} and \eqref{eq:non_y} to obtain
\begin{align}
    \left(\dfrac{B_{\rm ini}}{10^{-7}\,{\rm G}}\right)^{2}\left(\dfrac{\xi_{\rm M,ini}}{10^{-6}\,{\rm Mpc}}\right)<5\times 10^{-1},
    \label{eq:y_before_max}
\end{align}
for maximally helical fields and
\begin{align}
    \left(\dfrac{B_{\rm ini}}{10^{-7}\,{\rm G}}\right)^{4}\left(\dfrac{\xi_{\rm M,ini}}{10^{-6}\,{\rm Mpc}}\right)^5<2,
    \label{eq:y_before_non}
\end{align}
for non-helical fields.
These conditions define the left-side edges of the blue-shaded regions in Fig.~\ref{fig:constraint}.

For $z_{\rm ini}<z_{\rm BE}$, decay of the primordial magnetic fields does not generate $\mu$-distortions and is not constrained at all.
This explains that the $z_{\rm BE}$ lines define the right-side edges of the green-shaded regions in Fig.~\ref{fig:constraint}.

For $z_{\rm ini}<z_{\rm rec}$, generation of $y$-distortions by the decay of the primordial magnetic fields in the post-recombination epoch is not taken into account in our discussion.
Therefore, larger-scales beyond the $z_{\rm rec}$ lines in Fig.~\ref{fig:constraint} are not probed here.

\section{Discussion\label{sec:Discussion}}
\noindent
In this work, we update the constraint on primordial magnetic fields from the CMB spectral distortions, incorporating recent advancements in our understanding of MHD in the early universe.
In the previous studies such as \cite{2014JCAP...01..009K}, the constraint crucially depends on the spectral index of the energy power spectrum of the magnetic fields because they assumed that dissipation of linear MHD modes on small scales keeps long-range modes intact.
However, recently proposed was the idea that the conservation of either net magnetic helicity or the Hosking integral controls decay laws of primordial magnetic fields \cite{2021PhRvX..11d1005H,2023NatCo..14.7523H,2023PhLB..84338002U,2024arXiv240506194U}, regardless of the energy spectral index.
Based on this idea, we exclude the comoving strength and coherence length of primordial magnetic fields when they start decaying, $B_{\rm ini}$ and $\xi_{\rm M,ini}$, in the shaded regions in Fig.~\ref{fig:constraint}.
We emphasize that, with the formula recently developed, the coherence length of the excluded region turned out to be smaller than the previous studies due to the non-linear dynamics.

With the constraint on $B_{\rm ini}$ and $\xi_{\rm M, ini}$ approximated by Eqs.~\eqref{eq:oestimate} -- \eqref{eq:y_before_non}, we can obtain a constraint on magnetic fields at the recombination epoch, $B_0$ and $\xi_{{\rm M},0}$, which are to be compared with the observations in the late universe.
Note that here magnetic fields are approximated to evolve adiabatically after recombination until today.
Let us assume that primordial magnetic fields start decaying before $z_{\rm th}$, or equivalently, assume that they are generated in the regions above the apricot lines labeled $z_{\rm th}$ in Fig.~\ref{fig:constraint}.
The constraints from $\mu$-distortion, Eqs.~\eqref{eq:mu_before_max} and \eqref{eq:mu_before_non}, can be substituted into Eq.~\eqref{eq:cond_decaying} with $\tau=\tau_{\rm rec}$ to obtain
\begin{align}
    B_{\rm rec}&<3\times 10^{-9}\,{\rm G},&&{\text{(maximally helical)}}\label{eq:max_constraint}\\
    B_{\rm rec}&<7\times 10^{-10}\,{\rm G}.&&{\text{(non-helical)}}\label{eq:non_constraint}
\end{align}
These constraints are comparable to the ones from CMB anisotropy \cite{2013A&ARv..21...62D}, but the relevant scales are different.
While CMB anisotropy is sensitive to the scales beyond the Silk scale, $\gtrsim 10\,{\rm Mpc}$ at the recombination, our constraint corresponds to smaller scales, $\xi_{\rm M,rec}<10^{-6\text{--}5}\,{\rm Mpc}$, which may be compared with the constraints from the abundance of primordial black holes \cite{2020JCAP...05..039S,2024arXiv240519693K} and the magnetic reheating~\cite{2018MNRAS.474L..52S}.

However, it should be noted that the assumption of the decay timescale \eqref{eq:timescale_fast}, 
which makes the relevant magnetic coherence length shorter, is still under debate.
To discuss it, let us introduce a parameter $C_{\rm M}$ such that
\begin{align}
    v_{\rm A,rec}\xi^{-1}_{\rm M,rec}=C_{\rm M}\tau_{\rm rec}^{-1}. 
\end{align}
The difference in the decay timescale in the literature can be parameterized by the difference in the value of $C_\mathrm{M}$.
In the formulae we adopted~\cite{2021PhRvX..11d1005H,2023NatCo..14.7523H,2023PhLB..84338002U,2024arXiv240506194U}, it is taken as $C_{\rm M}=S_{\rm c}^{1/2}\sigma_{\rm rec}^{1/2}\eta_{\rm rec}^{1/2}\sim10^{3.5}$ (see Eq.~\eqref{eq:timescale_fast}).
Reference \cite{2004PhRvD..70l3003B} assumes the Alfv\'{e}n crossing time as the decay timescale to obtain $C'_{\rm M}=1$.
Reference \cite{2023NatCo..14.7523H} assumes $C''_{\rm M}=\tau^{1/2}_{\rm rec}/R^{1/2}_{\rm rec}l^{1/2}_{\gamma,{\rm rec}}\sim10$, by arguing that, even though magnetic reconnection takes place, photon drag controls the decay rate because the reconnection enters the so-called kinetic regime which is so fast that it does not limit the decay rate.
A numerical study \cite{Brandenburg:2024tyi} find $C'''_{\rm M}\sim\mathcal{O}(10)$, but this value is not to be directly compared with the discussions in the early universe because their setup does not include the photon drag, time-varying dissipation coefficients, nor a very large magnetic Prandtl number $\sigma\eta\gg 10^3$, which are all relevant in the recombination epoch.
If $C_\mathrm{M}$ is smaller, the coherence length of the excluded region moves to larger scales.
It remains unclear which value of $C_{\rm M}$ is correct, and numerical simulations with appropriate physics and relevant parameters are necessary to determine it.
We do not mean to exclude the possibility of smaller $C_\mathrm{M}$, but our treatment is the most conservative in the sense that $C'_{\rm M}, C''_{\rm M}, C'''_{\rm M}$ rule out much larger coherence lengths, reaching even $\gtrsim {\rm Mpc}$ scales.

Lastly, primordial magnetic fields may decay and generate CMB distortions even after the recombination epoch \cite{2017JCAP...12..011J}.
Although the analysis is not trivial because the electron and photon temperatures can be different in this era, it will extend the excluded region in Fig.~\ref{fig:constraint}.
From the spirit of the formulae recently developed~\cite{2021PhRvX..11d1005H,2023NatCo..14.7523H,2023PhLB..84338002U,2024arXiv240506194U}, the description of the evolution of magnetic fields after recombination should be also updated by taking into account the full non-linear effects, which is left for future study.

\section*{Acknowledgements}
\noindent
    We thank Eiichiro Komatsu for useful comments. 
    The work of FU was supported by JSPS KAKENHI Grant No.~23KJ0642.
    The work of KK was supported by the National Natural Science Foundation of China (NSFC) under Grant No. 12347103 and the JSPS KAKENHI Grant-in-Aid for Challenging Research (Exploratory) JP23K17687. The work of HT was supported by JSPS KAKENHI Grant No.~21K03533.

\appendix
\section{Electromagnetic units\label{qppx:emunits}}
\noindent
In this appendix, we review the relation between natural units to avoid confusion.
Similar discussion can be found in {\it e.g.}, footnote 5 of Ref.~\cite{2017JCAP...12..011J}.
In natural units, where $c=\hbar=1$, two conventions are commonly used for electromagnetic units: either setting $\epsilon_0\eqr1$ or $4\pi\epsilon_0\eqir1$.
The choice of convention varies across the literature.
Note that formulae such as $\rho_{\rm M}=\boldsymbol{B}^2/2\mu_0$ and $e^2=4\pi\epsilon_0\alpha\hbar c$ remain valid, but the specific values of $\epsilon_0$ and $\mu_0$ differ accordingly.

The difference leads to different relationships between Gauss (G) and GeV. 
From the definition of Gauss, $1 \mathrm{G}= 10^{-4} \,\mathrm{C}^{-1}\, \mathrm{kg}\, \mathrm{s}^{-1}$, we express
\begin{align}
    \mu_0^{-\frac{1}{2}}(\hbar c)^{\frac{3}{2}}
        =1.95\times 10^{-20}\,{\rm GeV}^2\,{\rm G}^{-1}
\end{align}
in the (non-natural) SI unit. 
This implies that
\begin{align}
    1\,{\rm G}
        \;\eqr1.95\times10^{-20}\,{\rm GeV}^2
\end{align}
in the natural Heaviside--Lorentz unit, where $\epsilon_0=1$ and $\mu_0=1$.  
On the other hand, we have
\begin{align}
    1\,{\rm G}
        \;\eqir6.92\times10^{-20}\,{\rm GeV}^2
\end{align}
in the natural cgs-Gauss unit, which adopts $4\pi\epsilon_0=1$ and $\mu_0=4\pi$.

Importantly, the expressions for the comoving electric conductivity and the comoving shear viscosity are different by powers of $4\pi$ depending on the choice of the unit convention. 
In this study, we estimate themvby the Spitzer theory \cite{1956pfig.book.....S}. 
In the natural Heaviside--Lorentz unit, they are given as
\begin{align}
    \sigma\,&\eqr\kappa_\sigma\dfrac{T_e^{\frac{3}{2}}}{zm_e^{\frac{1}{2}}e^{2}\ln\Lambda_{e{\rm i}}},\label{eq:sigma}\\
    \eta\,&\eqr\kappa_\eta\dfrac{T_{\rm i}^{\frac{5}{2}}m_{\rm i}^{\frac{1}{2}}}{z^3 e^{4}\rho_{\rm b}\ln\Lambda_{\rm ii}},\label{eq:eta}
\end{align}
where $m_e=0.5\,{\rm MeV}$ is the electron mass and $m_{\rm i}=1\,{\rm GeV}$ is the proton mass.
$\ln\Lambda_{e{\rm i}}\simeq\ln\Lambda_{{\rm ii}}\simeq20$ are the Coulomb logarithm, and we assume that the electron and ion temperatures coincide with the temperature of thermal plasma, $T_e\simeq T_\mathrm{i} \simeq T$ with $T=zT_0$.
The numerical prefactors $\kappa_\sigma$ and $\kappa_\eta$ depend on the approximation we adopt.
For the electric conductivity, under the Lorentz gas approximation, where the electron-electron collision is neglected, the coefficient is given by $\kappa'_\sigma=64\sqrt{2\pi}$ \cite{1950PhRv...80..230C,1956pfig.book.....S,2017ipp..book.....G}.
In some literature, an alternative value $\kappa''_\sigma=(4\pi)^2$ is also used, providing a close approximation to the Lorentz gas approximation~\cite{2023NatCo..14.7523H}.
Accounting for the corrections beyond the Lorentz gas approximation introduces a suppression factor $\sim0.6$~\cite{1953PhRv...89..977S}. 
Thus, in this work, we adopt an approximate value of $\kappa_\sigma\simeq 1\times10^2$. 
For the shear viscosity, we take $\kappa_\eta=4\sqrt{2}\pi/0.6\simeq 30$ \cite{2003phpl.book.....B}, based on the approximation that the ion velocity is $\sim\sqrt{2T_{\rm i}/m_{\rm i}}$. 
Alternatively, $\kappa'_\eta=(4\pi)^2$ \cite{2023NatCo..14.7523H}, may be obtained by dimensional analysis in the $4\pi\epsilon_0=1$ units.
The difference in this choice has caused the difference in the coherence length between our estimate and Ref.~\cite{2023NatCo..14.7523H}.

\section{Non-linearity estimate\label{appx:nonlinearity}}
\noindent
In this appendix, we review the way to the magnetic Reynolds number, following the approach outlined in Ref.~\cite{2024arXiv240506194U}, and clarify the condition when it is 
sufficiently large, Eq.~\eqref{eq:large_ReM}.

We first examine the breakdown of linearity under strong magnetic fields.
The magnetic Reynolds number could be estimated as \cite{2004PhRvD..70l3003B,2024arXiv240506194U}
\begin{align}
    {\rm Re}_{\rm M}
        &=\dfrac{\vert{\bm\nabla}\times({\bm v}\times{\bm B})\vert}{\vert \sigma^{-1} \nabla^2{\bm B}\vert}\notag\\
        &=\sigma v\xi=\dfrac{\sigma B^2\xi^2}{(\rho_{\rm fluid}+p_{\rm fluid})\eta}\quad\text{in linear regimes},
\end{align}
where we assume that, based on linearity, the magnetic field and the velocity field share the same coherence length, denoted as $\xi$, and that $v$ is related to $B$ by balancing the energy budget, expressed as $\vert({\bm\nabla}\times{\bm B})\times{\bm B}/(\rho_{\rm fluid}+p_{\rm fluid})\vert\sim\vert\eta\nabla^2{\bm v}\vert$ \cite{2004PhRvD..70l3003B}.
Then, the linearity condition ${\rm Re_{\rm M}}<1$ holds if
\begin{align}
    \left(\dfrac{B}{10^{-9}\,{\rm G}}\right)\left(\dfrac{\xi}{10^{-3}\,{\rm Mpc}}\right)
        <4\times 10^{-14}\,z^{-\frac{1}{2}}, 
    \label{eq:linearity}
\end{align}
where we have used Eqs.~\eqref{eq:sigma} and \eqref{eq:eta}.
However, this condition does not hold for the strong magnetic fields of our interest, $B\geq1\,{\rm nG}$ and $\xi_{\rm M}\geq10^{-9}\,{\rm Mpc}$, implying that the magnetic field is in a non-linear regime.\footnote{A subtlety is that if we assume equipartition $v\sim B/\sqrt{\rho_\mathrm{fluid}}$, the linearity condition becomes $B\xi<\sigma^{-1}$ and is satisfied with the relevant magnetic fields, $B\geq1\,{\rm nG}$ and $\xi_{\rm M}\geq10^{-9}\,{\rm Mpc}$. 
However, equipartition seems specific to the kinetically dominated and non-helical initial conditions \cite{2017PhRvD..96l3528B}, where the dissipation of the velocity field governs the decay dynamics and magnetic field is just enhanced at small scales \cite{2021PhRvX..11d1005H,2024arXiv240506194U}.
We believe that this estimate does not apply to the magnetically dominated fluid relevant to our study.}

In the non-linear regime, the Lundquist number also quantifies non-linearity.
In the context of magnetic reconnection, it can be defined as \cite{2022JPlPh..88e1501S,2023NatCo..14.7523H,2024arXiv240506194U}
\begin{align}
    S:=\sigma v_{\rm out}\xi_{\rm M},
\end{align}
where $v_{\rm out}$ is the velocity of the outflow from the current sheet.
In the original Sweet--Parker model \cite{1958IAUS....6..123S,1957JGR....62..509P}, the outflow velocity is identified with the Alfv\'{e}n velocity, $v_{\rm A}=B/\sqrt{\rho_{\rm fluid}+p_{\rm fluid}}$, which reproduces the standard definition of the Lundquist number.
However, the presence of finite viscosity dissipates the outflow inside the current sheets and suppresses the estimate by a factor $1/\sqrt{1+{\rm Pr}_{\rm M}}$, where ${\rm Pr}_{\rm M}:=\sigma\eta$ is the magnetic Prandtl number \cite{1984PhFl...27..137P}.
Since ${\rm Pr}_{\rm M}\gg1$ in the early universe before the recombination \cite{2004PhRvD..70l3003B,2011PhPl...18k1207J,2024arXiv240506194U}, we approximate
\begin{align}
    v_{\rm out}=B/\sqrt{(\rho_{\rm fluid}+p_{\rm fluid})\sigma\eta}.
\end{align}
Theoretical models of magnetic reconnection \cite{1957JGR....62..509P,1958IAUS....6..123S,2005PhRvL..95w5003L,2007PhPl...14j0703L} predict a general relation between the magnetic Reynolds number and the Lundquist number \cite{2024arXiv240506194U},
\begin{align}
    {\rm Re}_{\rm M}
        &=\dfrac{\vert{\bm\nabla}\times({\bm v}\times{\bm B})\vert}{\vert \sigma^{-1} \nabla^2{\bm B}\vert}\notag\\
        &=\dfrac{\sigma v\xi^2_{\rm M}}{\xi_{\rm K}}=S^{\frac{5}{4}}\quad\text{in reconnection regimes},
\end{align}
where we estimated $\boldsymbol{\nabla}\sim\xi_{\rm K}^{-1}$ in the numerator because the large aspect ratio of the current sheets in the reconnection regimes implies $\xi_{\rm K}\ll\xi_{\rm M}$.
In the last equality, we have used general relations, $v^2=(\xi_{\rm K}/\xi_{\rm M})v_{\rm out}^2$ and $v_{\rm out}=\xi_{\rm M}/(\sigma\xi_{\rm K}^2)$, adopted in the models of magnetic reconnection~\cite{1957JGR....62..509P,1958IAUS....6..123S,2005PhRvL..95w5003L,2007PhPl...14j0703L}.
The Lundquist number is larger than the critical value $S_{\rm c}\sim10^4$ \cite{2011PhPl...18k1207J,1986PhFl...29.1520B,2005PhRvL..95w5003L,2007PhPl...14j0703L} if
\begin{align}
    \left(\dfrac{B}{10^{-9}\,{\rm G}}\right)\left(\dfrac{\xi_{\rm M}}{10^{-3}\,{\rm Mpc}}\right)
        >1\times 10^{-11}\,z^{-\frac{1}{2}}.
    \label{eq:nonlinearity}
\end{align}
Once more, this condition is met for the magnetic fields of our interest, $B\geq1\,{\rm nG}$ and $\xi_{\rm M}\geq10^{-9}\,{\rm Mpc}$, implying that the highly non-linear fast magnetic reconnection is taking place \cite{2023NatCo..14.7523H}.

\section{The Hosking integral}\label{appx:HI}
\noindent
In this appendix, we review the Hosking integral.
The Hosking integral is defined as \cite{2021PhRvX..11d1005H}
\begin{align}
    I_{\rm H}
        :=\int d^3r\,\left\langle h(\boldsymbol x)h(\boldsymbol x+\boldsymbol r)\right\rangle,
\end{align}
where $h=\boldsymbol A\cdot\boldsymbol B$.
Although $h(\boldsymbol{x})$ is a gauge-dependent quantity, $I_{\rm H}$ is gauge-independent, finite, and conserved for a non-helical magnetic field if one takes the integration volume much larger than the coherence length $\xi_{\rm M}$ \cite{2022JPlPh..88f9002Z}.

To evaluate $I_{\rm H}$, we may approximate
\begin{align}
    \left\langle h(\boldsymbol x)h(\boldsymbol x+\boldsymbol r)\right\rangle
        \simeq\begin{cases}
            \langle h^2\rangle&r\ll\xi_{\rm M},\\
            0&r\gg\xi_{\rm M},
        \end{cases}
\end{align}
where $\langle h^2\rangle\sim B^4\xi_{\rm M}^2$ if we assume Gaussianity of the probability distribution of the magnetic field.
Then, we obtain a proportionality
\begin{align}
    I_{\rm H}
        \sim B^4\xi_{\rm M}^5,
\end{align}
where a nearly time-independent numerical coefficient, determined by the shape of the magnetic power spectrum, is included implicitly.
Note that non-Gaussianity can introduce an additional time-dependent factor, see Ref.~\cite{2022JPlPh..88f9002Z}.
However, since the behavior of this factor is poorly understood and since the numerical results agree well with the conservation law \eqref{eq:non-}, we just adopt the Gaussian approximation.

\bibliographystyle{elsarticle-num}
\bibliography{revise}

\begin{thebibliography}{10}
\expandafter\ifx\csname url\endcsname\relax
  \def\url#1{\texttt{#1}}\fi
\expandafter\ifx\csname urlprefix\endcsname\relax\def\urlprefix{URL }\fi
\expandafter\ifx\csname href\endcsname\relax
  \def\href#1#2{#2} \def\path#1{#1}\fi

\bibitem{1998PhRvD..57.3264J}
K.~{Jedamzik}, V.~{Katalini{\'c}}, A.~V. {Olinto}, {Damping of cosmic magnetic fields}, \prd 57~(6) (1998) 3264--3284.
\newblock \href {http://arxiv.org/abs/astro-ph/9606080} {\path{arXiv:astro-ph/9606080}}, \href {https://doi.org/10.1103/PhysRevD.57.3264} {\path{doi:10.1103/PhysRevD.57.3264}}.

\bibitem{1988PhRvD..37.2743T}
M.~S. {Turner}, L.~M. {Widrow}, Inflation-produced, large-scale magnetic fields, \prd 37~(10) (1988) 2743--2754.
\newblock \href {https://doi.org/10.1103/PhysRevD.37.2743} {\path{doi:10.1103/PhysRevD.37.2743}}.

\bibitem{1992ApJ...391L...1R}
B.~{Ratra}, Cosmological ``seed'' magnetic field from inflation, \apjl 391 (1992) L1.
\newblock \href {https://doi.org/10.1086/186384} {\path{doi:10.1086/186384}}.

\bibitem{1992PhRvD..46.5346G}
W.~D. {Garretson}, G.~B. {Field}, S.~M. {Carroll}, Primordial magnetic fields from pseudo goldstone bosons, \prd 46~(12) (1992) 5346--5351.
\newblock \href {http://arxiv.org/abs/hep-ph/9209238} {\path{arXiv:hep-ph/9209238}}, \href {https://doi.org/10.1103/PhysRevD.46.5346} {\path{doi:10.1103/PhysRevD.46.5346}}.

\bibitem{PhysRevLett.51.1488}
C.~J. Hogan, Magnetohydrodynamic effects of a first-order cosmological phase transition, Phys. Rev. Lett. 51 (1983) 1488--1491.
\newblock \href {https://doi.org/10.1103/PhysRevLett.51.1488} {\path{doi:10.1103/PhysRevLett.51.1488}}.

\bibitem{1989ApJ...344L..49Q}
J.~M. {Quashnock}, A.~{Loeb}, D.~N. {Spergel}, Magnetic field generation during the cosmological qcd phase transition, \apjl 344 (1989) L49.
\newblock \href {https://doi.org/10.1086/185528} {\path{doi:10.1086/185528}}.

\bibitem{PhysRevD.50.2421}
B.~Cheng, A.~V. Olinto, Primordial magnetic fields generated in the quark-hadron transition, Phys. Rev. D 50 (1994) 2421--2424.
\newblock \href {https://doi.org/10.1103/PhysRevD.50.2421} {\path{doi:10.1103/PhysRevD.50.2421}}.

\bibitem{1991PhLB..265..258V}
T.~{Vachaspati}, Magnetic fields from cosmological phase transitions, Physics Letters B 265~(3-4) (1991) 258--261.
\newblock \href {https://doi.org/10.1016/0370-2693(91)90051-Q} {\path{doi:10.1016/0370-2693(91)90051-Q}}.

\bibitem{2010Sci...328...73N}
A.~{Neronov}, I.~{Vovk}, {Evidence for Strong Extragalactic Magnetic Fields from Fermi Observations of TeV Blazars}, Science 328~(5974) (2010) 73.
\newblock \href {http://arxiv.org/abs/1006.3504} {\path{arXiv:1006.3504}}, \href {https://doi.org/10.1126/science.1184192} {\path{doi:10.1126/science.1184192}}.

\bibitem{2010MNRAS.406L..70T}
F.~{Tavecchio}, G.~{Ghisellini}, L.~{Foschini}, G.~{Bonnoli}, G.~{Ghirlanda}, P.~{Coppi}, {The intergalactic magnetic field constrained by Fermi/Large Area Telescope observations of the TeV blazar 1ES0229+200}, \mnras 406~(1) (2010) L70--L74.
\newblock \href {http://arxiv.org/abs/1004.1329} {\path{arXiv:1004.1329}}, \href {https://doi.org/10.1111/j.1745-3933.2010.00884.x} {\path{doi:10.1111/j.1745-3933.2010.00884.x}}.

\bibitem{2011ApJ...727L...4D}
K.~{Dolag}, M.~{Kachelriess}, S.~{Ostapchenko}, R.~{Tom{\`a}s}, {Lower Limit on the Strength and Filling Factor of Extragalactic Magnetic Fields}, \apjl 727~(1) (2011) L4.
\newblock \href {http://arxiv.org/abs/1009.1782} {\path{arXiv:1009.1782}}, \href {https://doi.org/10.1088/2041-8205/727/1/L4} {\path{doi:10.1088/2041-8205/727/1/L4}}.

\bibitem{2018ApJS..237...32A}
M.~{Ackermann}, et~al., {The Search for Spatial Extension in High-latitude Sources Detected by the Fermi Large Area Telescope}, \apjs 237~(2) (2018) 32.
\newblock \href {http://arxiv.org/abs/1804.08035} {\path{arXiv:1804.08035}}, \href {https://doi.org/10.3847/1538-4365/aacdf7} {\path{doi:10.3847/1538-4365/aacdf7}}.

\bibitem{2023A&A...670A.145A}
V.~A. {Acciari}, et~al., {A lower bound on intergalactic magnetic fields from time variability of 1ES 0229+200 from MAGIC and Fermi/LAT observations}, \aap 670 (2023) A145.
\newblock \href {http://arxiv.org/abs/2210.03321} {\path{arXiv:2210.03321}}, \href {https://doi.org/10.1051/0004-6361/202244126} {\path{doi:10.1051/0004-6361/202244126}}.

\bibitem{2024MNRAS.527L..95D}
T.~A. {Dzhatdoev}, E.~I. {Podlesnyi}, G.~I. {Rubtsov}, {First constraints on the strength of the extragalactic magnetic field from {\ensuremath{\gamma}}-ray observations of GRB 221009A}, \mnras 527~(1) (2024) L95--L102.
\newblock \href {http://arxiv.org/abs/2306.05347} {\path{arXiv:2306.05347}}, \href {https://doi.org/10.1093/mnrasl/slad142} {\path{doi:10.1093/mnrasl/slad142}}.

\bibitem{2020JCAP...05..039S}
S.~{Saga}, H.~{Tashiro}, S.~{Yokoyama}, {Limits on primordial magnetic fields from primordial black hole abundance}, \jcap 2020~(5) (2020) 039.
\newblock \href {http://arxiv.org/abs/2002.01286} {\path{arXiv:2002.01286}}, \href {https://doi.org/10.1088/1475-7516/2020/05/039} {\path{doi:10.1088/1475-7516/2020/05/039}}.

\bibitem{2024arXiv240519693K}
A.~{Kushwaha}, T.~{Suyama}, {Constraining small-scale primordial magnetic fields from the abundance of primordial black holes}, arXiv e-prints (2024) arXiv:2405.19693\href {http://arxiv.org/abs/2405.19693} {\path{arXiv:2405.19693}}, \href {https://doi.org/10.48550/arXiv.2405.19693} {\path{doi:10.48550/arXiv.2405.19693}}.

\bibitem{2018MNRAS.474L..52S}
S.~{Saga}, H.~{Tashiro}, S.~{Yokoyama}, {Magnetic reheating}, \mnras 474~(1) (2018) L52--L55.
\newblock \href {http://arxiv.org/abs/1708.08225} {\path{arXiv:1708.08225}}, \href {https://doi.org/10.1093/mnrasl/slx195} {\path{doi:10.1093/mnrasl/slx195}}.

\bibitem{1998PhRvD..57.2186G}
M.~{Giovannini}, M.~E. {Shaposhnikov}, {Primordial hypermagnetic fields and the triangle anomaly}, \prd 57~(4) (1998) 2186--2206.
\newblock \href {http://arxiv.org/abs/hep-ph/9710234} {\path{arXiv:hep-ph/9710234}}, \href {https://doi.org/10.1103/PhysRevD.57.2186} {\path{doi:10.1103/PhysRevD.57.2186}}.

\bibitem{2016PhRvD..93h3520F}
T.~{Fujita}, K.~{Kamada}, {Large-scale magnetic fields can explain the baryon asymmetry of the Universe}, \prd 93~(8) (2016) 083520.
\newblock \href {http://arxiv.org/abs/1602.02109} {\path{arXiv:1602.02109}}, \href {https://doi.org/10.1103/PhysRevD.93.083520} {\path{doi:10.1103/PhysRevD.93.083520}}.

\bibitem{2021JCAP...04..034K}
K.~{Kamada}, F.~{Uchida}, J.~{Yokoyama}, {Baryon isocurvature constraints on the primordial hypermagnetic fields}, \jcap 2021~(4) (2021) 034.
\newblock \href {http://arxiv.org/abs/2012.14435} {\path{arXiv:2012.14435}}, \href {https://doi.org/10.1088/1475-7516/2021/04/034} {\path{doi:10.1088/1475-7516/2021/04/034}}.

\bibitem{1969Natur.223..938G}
G.~{Greenstein}, {Primordial Helium Production in ``Magnetic'' Cosmologies}, \nat 223~(5209) (1969) 938--939.
\newblock \href {https://doi.org/10.1038/223938b0} {\path{doi:10.1038/223938b0}}.

\bibitem{2018PhRvD..98h3518S}
S.~{Saga}, H.~{Tashiro}, S.~{Yokoyama}, {Limits on primordial magnetic fields from direct detection experiments of gravitational wave background}, \prd 98~(8) (2018) 083518.
\newblock \href {http://arxiv.org/abs/1807.00561} {\path{arXiv:1807.00561}}, \href {https://doi.org/10.1103/PhysRevD.98.083518} {\path{doi:10.1103/PhysRevD.98.083518}}.

\bibitem{2022PhRvD.105l3502R}
A.~{Roper Pol}, C.~{Caprini}, A.~{Neronov}, D.~{Semikoz}, {Gravitational wave signal from primordial magnetic fields in the Pulsar Timing Array frequency band}, \prd 105~(12) (2022) 123502.
\newblock \href {http://arxiv.org/abs/2201.05630} {\path{arXiv:2201.05630}}, \href {https://doi.org/10.1103/PhysRevD.105.123502} {\path{doi:10.1103/PhysRevD.105.123502}}.

\bibitem{2016A&A...594A..19P}
{Planck Collaboration}, {Planck 2015 results. XIX. Constraints on primordial magnetic fields}, \aap 594 (2016) A19.
\newblock \href {http://arxiv.org/abs/1502.01594} {\path{arXiv:1502.01594}}, \href {https://doi.org/10.1051/0004-6361/201525821} {\path{doi:10.1051/0004-6361/201525821}}.

\bibitem{2024JCAP...07..086P}
{LiteBIRD Collaboration}, {LiteBIRD science goals and forecasts: primordial magnetic fields}, \jcap 2024~(7) (2024) 086.
\newblock \href {http://arxiv.org/abs/2403.16763} {\path{arXiv:2403.16763}}, \href {https://doi.org/10.1088/1475-7516/2024/07/086} {\path{doi:10.1088/1475-7516/2024/07/086}}.

\bibitem{1996ApJ...468...28K}
E.-J. {Kim}, A.~V. {Olinto}, R.~{Rosner}, {Generation of Density Perturbations by Primordial Magnetic Fields}, \apj 468 (1996) 28.
\newblock \href {http://arxiv.org/abs/astro-ph/9412070} {\path{arXiv:astro-ph/9412070}}, \href {https://doi.org/10.1086/177667} {\path{doi:10.1086/177667}}.

\bibitem{2024PhRvD.109d3520S}
S.~{Saga}, M.~{Shiraishi}, K.~{Akitsu}, T.~{Okumura}, {Imprints of primordial magnetic fields on intrinsic alignments of galaxies}, \prd 109~(4) (2024) 043520.
\newblock \href {http://arxiv.org/abs/2312.16316} {\path{arXiv:2312.16316}}, \href {https://doi.org/10.1103/PhysRevD.109.043520} {\path{doi:10.1103/PhysRevD.109.043520}}.

\bibitem{2023PASJ...75S.154M}
T.~{Minoda}, S.~{Saga}, T.~{Takahashi}, H.~{Tashiro}, D.~{Yamauchi}, S.~{Yokoyama}, S.~{Yoshiura}, {Probing the primordial Universe with 21 cm line from cosmic dawn/epoch of reionization}, \pasj 75 (2023) S154--S180.
\newblock \href {http://arxiv.org/abs/2303.07604} {\path{arXiv:2303.07604}}, \href {https://doi.org/10.1093/pasj/psac015} {\path{doi:10.1093/pasj/psac015}}.

\bibitem{2024ApJ...972..117Z}
Q.~{Zhang}, S.~{Li}, X.-H. {Tan}, J.-Q. {Xia}, {Constraints on Primordial Magnetic Fields from High Redshift Stellar Mass Density}, \apj 972~(1) (2024) 117.
\newblock \href {http://arxiv.org/abs/2408.03584} {\path{arXiv:2408.03584}}, \href {https://doi.org/10.3847/1538-4357/ad685e} {\path{doi:10.3847/1538-4357/ad685e}}.

\bibitem{1970Ap&SS...7....3S}
R.~A. {Sunyaev}, Y.~B. {Zeldovich}, {Small-Scale Fluctuations of Relic Radiation}, \apss 7~(1) (1970) 3--19.
\newblock \href {https://doi.org/10.1007/BF00653471} {\path{doi:10.1007/BF00653471}}.

\bibitem{1980ARA&A..18..537S}
R.~A. {Sunyaev}, I.~B. {Zeldovich}, {Microwave background radiation as a probe of the contemporary structure and history of the universe}, \araa 18 (1980) 537--560.
\newblock \href {https://doi.org/10.1146/annurev.aa.18.090180.002541} {\path{doi:10.1146/annurev.aa.18.090180.002541}}.

\bibitem{1995A&A...303..323B}
C.~{Burigana}, G.~{de Zotti}, L.~{Danese}, {Analytical description of spectral distortions of the cosmic microwave background.}, \aap 303 (1995) 323.

\bibitem{Jedamzik:1999bm}
K.~Jedamzik, V.~Katalinic, A.~V. Olinto, {A Limit on primordial small scale magnetic fields from CMB distortions}, Phys. Rev. Lett. 85 (2000) 700--703.
\newblock \href {http://arxiv.org/abs/astro-ph/9911100} {\path{arXiv:astro-ph/9911100}}, \href {https://doi.org/10.1103/PhysRevLett.85.700} {\path{doi:10.1103/PhysRevLett.85.700}}.

\bibitem{2005MNRAS.356..778S}
S.~K. {Sethi}, K.~{Subramanian}, {Primordial magnetic fields in the post-recombination era and early reionization}, \mnras 356~(2) (2005) 778--788.
\newblock \href {http://arxiv.org/abs/astro-ph/0405413} {\path{arXiv:astro-ph/0405413}}, \href {https://doi.org/10.1111/j.1365-2966.2004.08520.x} {\path{doi:10.1111/j.1365-2966.2004.08520.x}}.

\bibitem{2014JCAP...01..009K}
K.~E. {Kunze}, E.~{Komatsu}, {Constraining primordial magnetic fields with distortions of the black-body spectrum of the cosmic microwave background: pre- and post-decoupling contributions}, \jcap 2014~(1) (2014) 009.
\newblock \href {http://arxiv.org/abs/1309.7994} {\path{arXiv:1309.7994}}, \href {https://doi.org/10.1088/1475-7516/2014/01/009} {\path{doi:10.1088/1475-7516/2014/01/009}}.

\bibitem{2015PhRvD..92l3004W}
J.~M. {Wagstaff}, R.~{Banerjee}, {CMB spectral distortions from the decay of causally generated magnetic fields}, \prd 92~(12) (2015) 123004.
\newblock \href {http://arxiv.org/abs/1508.01683} {\path{arXiv:1508.01683}}, \href {https://doi.org/10.1103/PhysRevD.92.123004} {\path{doi:10.1103/PhysRevD.92.123004}}.

\bibitem{2019MNRAS.490.4419S}
S.~{Saga}, A.~{Ota}, H.~{Tashiro}, S.~{Yokoyama}, {Secondary CMB temperature anisotropies from magnetic reheating}, \mnras 490~(3) (2019) 4419--4427.
\newblock \href {http://arxiv.org/abs/1904.09121} {\path{arXiv:1904.09121}}, \href {https://doi.org/10.1093/mnras/stz2882} {\path{doi:10.1093/mnras/stz2882}}.

\bibitem{1966ApJ...145..560W}
R.~{Weymann}, {The Energy Spectrum of Radiation in the Expanding Universe}, \apj 145 (1966) 560.
\newblock \href {https://doi.org/10.1086/148795} {\path{doi:10.1086/148795}}.

\bibitem{1969Ap&SS...4..301Z}
Y.~B. {Zeldovich}, R.~A. {Sunyaev}, {The Interaction of Matter and Radiation in a Hot-Model Universe}, \apss 4~(3) (1969) 301--316.
\newblock \href {https://doi.org/10.1007/BF00661821} {\path{doi:10.1007/BF00661821}}.

\bibitem{1970Ap&SS...7...20S}
R.~A. {Sunyaev}, Y.~B. {Zeldovich}, {The interaction of matter and radiation in the hot model of the Universe, II}, \apss 7~(1) (1970) 20--30.
\newblock \href {https://doi.org/10.1007/BF00653472} {\path{doi:10.1007/BF00653472}}.

\bibitem{1982A&A...107...39D}
L.~{Danese}, G.~{de Zotti}, {Double Compton process and the spectrum of the microwave background}, \aap 107~(1) (1982) 39--42.

\bibitem{1993PhRvD..48..485H}
W.~{Hu}, J.~{Silk}, {Thermalization and spectral distortions of the cosmic background radiation}, \prd 48~(2) (1993) 485--502.
\newblock \href {https://doi.org/10.1103/PhysRevD.48.485} {\path{doi:10.1103/PhysRevD.48.485}}.

\bibitem{2012MNRAS.425.1129C}
J.~{Chluba}, R.~{Khatri}, R.~A. {Sunyaev}, {CMB at 2 {\texttimes} 2 order: the dissipation of primordial acoustic waves and the observable part of the associated energy release}, \mnras 425~(2) (2012) 1129--1169.
\newblock \href {http://arxiv.org/abs/1202.0057} {\path{arXiv:1202.0057}}, \href {https://doi.org/10.1111/j.1365-2966.2012.21474.x} {\path{doi:10.1111/j.1365-2966.2012.21474.x}}.

\bibitem{2012A&A...543A.136K}
R.~{Khatri}, R.~A. {Sunyaev}, J.~{Chluba}, {Mixing of blackbodies: entropy production and dissipation of sound waves in the early Universe}, \aap 543 (2012) A136.
\newblock \href {http://arxiv.org/abs/1205.2871} {\path{arXiv:1205.2871}}, \href {https://doi.org/10.1051/0004-6361/201219590} {\path{doi:10.1051/0004-6361/201219590}}.

\bibitem{2013MNRAS.434..352C}
J.~{Chluba}, {Green's function of the cosmological thermalization problem}, \mnras 434~(1) (2013) 352--357.
\newblock \href {http://arxiv.org/abs/1304.6120} {\path{arXiv:1304.6120}}, \href {https://doi.org/10.1093/mnras/stt1025} {\path{doi:10.1093/mnras/stt1025}}.

\bibitem{1998PhRvD..58h3502S}
K.~{Subramanian}, J.~D. {Barrow}, {Magnetohydrodynamics in the early universe and the damping of nonlinear Alfv{\'e}n waves}, \prd 58~(8) (1998) 083502.
\newblock \href {http://arxiv.org/abs/astro-ph/9712083} {\path{arXiv:astro-ph/9712083}}, \href {https://doi.org/10.1103/PhysRevD.58.083502} {\path{doi:10.1103/PhysRevD.58.083502}}.

\bibitem{2004PhRvD..70l3003B}
R.~{Banerjee}, K.~{Jedamzik}, {Evolution of cosmic magnetic fields: From the very early Universe, to recombination, to the present}, \prd 70~(12) (2004) 123003.
\newblock \href {http://arxiv.org/abs/astro-ph/0410032} {\path{arXiv:astro-ph/0410032}}, \href {https://doi.org/10.1103/PhysRevD.70.123003} {\path{doi:10.1103/PhysRevD.70.123003}}.

\bibitem{2014ApJ...794L..26Z}
J.~{Zrake}, {Inverse Cascade of Nonhelical Magnetic Turbulence in a Relativistic Fluid}, \apjl 794~(2) (2014) L26.
\newblock \href {http://arxiv.org/abs/1407.5626} {\path{arXiv:1407.5626}}, \href {https://doi.org/10.1088/2041-8205/794/2/L26} {\path{doi:10.1088/2041-8205/794/2/L26}}.

\bibitem{2015PhRvL.114g5001B}
A.~{Brandenburg}, T.~{Kahniashvili}, A.~G. {Tevzadze}, {Nonhelical Inverse Transfer of a Decaying Turbulent Magnetic Field}, \prl 114~(7) (2015) 075001.
\newblock \href {http://arxiv.org/abs/1404.2238} {\path{arXiv:1404.2238}}, \href {https://doi.org/10.1103/PhysRevLett.114.075001} {\path{doi:10.1103/PhysRevLett.114.075001}}.

\bibitem{2001PhRvE..64e6405C}
M.~{Christensson}, M.~{Hindmarsh}, A.~{Brandenburg}, {Inverse cascade in decaying three-dimensional magnetohydrodynamic turbulence}, \pre 64~(5) (2001) 056405.
\newblock \href {http://arxiv.org/abs/astro-ph/0011321} {\path{arXiv:astro-ph/0011321}}, \href {https://doi.org/10.1103/PhysRevE.64.056405} {\path{doi:10.1103/PhysRevE.64.056405}}.

\bibitem{Brandenburg:2016odr}
A.~Brandenburg, T.~Kahniashvili, Classes of hydrodynamic and magnetohydrodynamic turbulent decay, Phys. Rev. Lett. 118~(5) (2017) 055102.
\newblock \href {http://arxiv.org/abs/1607.01360} {\path{arXiv:1607.01360}}, \href {https://doi.org/10.1103/PhysRevLett.118.055102} {\path{doi:10.1103/PhysRevLett.118.055102}}.

\bibitem{Brandenburg+17}
A.~Brandenburg, T.~Kahniashvili, S.~Mandal, A.~R. Pol, A.~G. Tevzadze, T.~Vachaspati, Evolution of hydromagnetic turbulence from the electroweak phase transition, Physical Review D 96~(12) (2017) 123528.

\bibitem{2017PhRvE..96e3105R}
J.~{Reppin}, R.~{Banerjee}, {Nonhelical turbulence and the inverse transfer of energy: A parameter study}, \pre 96~(5) (2017) 053105.
\newblock \href {http://arxiv.org/abs/1708.07717} {\path{arXiv:1708.07717}}, \href {https://doi.org/10.1103/PhysRevE.96.053105} {\path{doi:10.1103/PhysRevE.96.053105}}.

\bibitem{2017MNRAS.472.1628P}
K.~{Park}, {On the inverse transfer of (non-)helical magnetic energy in a decaying magnetohydrodynamic turbulence}, \mnras 472~(2) (2017) 1628--1640.
\newblock \href {http://arxiv.org/abs/1709.06526} {\path{arXiv:1709.06526}}, \href {https://doi.org/10.1093/mnras/stx1981} {\path{doi:10.1093/mnras/stx1981}}.

\bibitem{2021PhRvX..11d1005H}
D.~N. {Hosking}, A.~A. {Schekochihin}, {Reconnection-Controlled Decay of Magnetohydrodynamic Turbulence and the Role of Invariants}, Physical Review X 11~(4) (2021) 041005.
\newblock \href {http://arxiv.org/abs/2012.01393} {\path{arXiv:2012.01393}}, \href {https://doi.org/10.1103/PhysRevX.11.041005} {\path{doi:10.1103/PhysRevX.11.041005}}.

\bibitem{2023NatCo..14.7523H}
D.~N. {Hosking}, A.~A. {Schekochihin}, {Cosmic-void observations reconciled with primordial magnetogenesis}, Nature Communications 14 (2023) 7523.
\newblock \href {http://arxiv.org/abs/2203.03573} {\path{arXiv:2203.03573}}, \href {https://doi.org/10.1038/s41467-023-43258-3} {\path{doi:10.1038/s41467-023-43258-3}}.

\bibitem{2023PhLB..84338002U}
F.~{Uchida}, M.~{Fujiwara}, K.~{Kamada}, J.~{Yokoyama}, {New description of the scaling evolution of the cosmological magneto-hydrodynamic system}, Physics Letters B 843 (2023) 138002.
\newblock \href {http://arxiv.org/abs/2212.14355} {\path{arXiv:2212.14355}}, \href {https://doi.org/10.1016/j.physletb.2023.138002} {\path{doi:10.1016/j.physletb.2023.138002}}.

\bibitem{2024arXiv240506194U}
F.~{Uchida}, M.~{Fujiwara}, K.~{Kamada}, J.~{Yokoyama}, {New comprehensive description of the scaling evolution of the cosmological magneto-hydrodynamic system}, arXiv e-prints (2024) arXiv:2405.06194\href {http://arxiv.org/abs/2405.06194} {\path{arXiv:2405.06194}}, \href {https://doi.org/10.48550/arXiv.2405.06194} {\path{doi:10.48550/arXiv.2405.06194}}.

\bibitem{2013PhRvD..87h3007K}
T.~{Kahniashvili}, A.~G. {Tevzadze}, A.~{Brandenburg}, A.~{Neronov}, {Evolution of primordial magnetic fields from phase transitions}, \prd 87~(8) (2013) 083007.
\newblock \href {http://arxiv.org/abs/1212.0596} {\path{arXiv:1212.0596}}, \href {https://doi.org/10.1103/PhysRevD.87.083007} {\path{doi:10.1103/PhysRevD.87.083007}}.

\bibitem{2021MNRAS.501.3074B}
P.~{Bhat}, M.~{Zhou}, N.~F. {Loureiro}, {Inverse energy transfer in decaying, three-dimensional, non-helical magnetic turbulence due to magnetic reconnection}, \mnras 501~(2) (2021) 3074--3087.
\newblock \href {http://arxiv.org/abs/2007.07325} {\path{arXiv:2007.07325}}, \href {https://doi.org/10.1093/mnras/staa3849} {\path{doi:10.1093/mnras/staa3849}}.

\bibitem{2024OJAp....7E..75D}
S.~{Dwivedi}, C.~{Anandavijayan}, P.~{Bhat}, {Quasi-two-dimensionality of three-dimensional, magnetically dominated, decaying turbulence}, The Open Journal of Astrophysics 7 (2024) 75.
\newblock \href {http://arxiv.org/abs/2401.01965} {\path{arXiv:2401.01965}}, \href {https://doi.org/10.33232/001c.122855} {\path{doi:10.33232/001c.122855}}.

\bibitem{2020A&A...641A...6P}
{Planck Collaboration}, {Planck 2018 results. VI. Cosmological parameters}, \aap 641 (2020) A6.
\newblock \href {http://arxiv.org/abs/1807.06209} {\path{arXiv:1807.06209}}, \href {https://doi.org/10.1051/0004-6361/201833910} {\path{doi:10.1051/0004-6361/201833910}}.

\bibitem{1968ApJ...151..459S}
J.~{Silk}, {Cosmic Black-Body Radiation and Galaxy Formation}, \apj 151 (1968) 459.
\newblock \href {https://doi.org/10.1086/149449} {\path{doi:10.1086/149449}}.

\bibitem{2017PhRvD..96l3528B}
A.~{Brandenburg}, T.~{Kahniashvili}, S.~{Mandal}, A.~R. {Pol}, A.~G. {Tevzadze}, T.~{Vachaspati}, {Evolution of hydromagnetic turbulence from the electroweak phase transition}, \prd 96~(12) (2017) 123528.
\newblock \href {http://arxiv.org/abs/1711.03804} {\path{arXiv:1711.03804}}, \href {https://doi.org/10.1103/PhysRevD.96.123528} {\path{doi:10.1103/PhysRevD.96.123528}}.

\bibitem{1957JGR....62..509P}
E.~N. {Parker}, {Sweet's Mechanism for Merging Magnetic Fields in Conducting Fluids}, \jgr 62~(4) (1957) 509--520.
\newblock \href {https://doi.org/10.1029/JZ062i004p00509} {\path{doi:10.1029/JZ062i004p00509}}.

\bibitem{1958IAUS....6..123S}
P.~A. {Sweet}, {The Neutral Point Theory of Solar Flares}, in: B.~{Lehnert} (Ed.), Electromagnetic Phenomena in Cosmical Physics, Vol.~6, 1958, p. 123.

\bibitem{1984PhFl...27..137P}
W.~{Park}, D.~A. {Monticello}, R.~B. {White}, {Reconnection rates of magnetic fields including the effects of viscosity}, Physics of Fluids 27~(1) (1984) 137--149.
\newblock \href {https://doi.org/10.1063/1.864502} {\path{doi:10.1063/1.864502}}.

\bibitem{2011PhPl...18k1207J}
H.~{Ji}, W.~{Daughton}, {Phase diagram for magnetic reconnection in heliophysical, astrophysical, and laboratory plasmas}, Physics of Plasmas 18~(11) (2011) 111207--111207.
\newblock \href {http://arxiv.org/abs/1109.0756} {\path{arXiv:1109.0756}}, \href {https://doi.org/10.1063/1.3647505} {\path{doi:10.1063/1.3647505}}.

\bibitem{1986PhFl...29.1520B}
D.~{Biskamp}, {Magnetic reconnection via current sheets}, Physics of Fluids 29~(5) (1986) 1520--1531.
\newblock \href {https://doi.org/10.1063/1.865670} {\path{doi:10.1063/1.865670}}.

\bibitem{2005PhRvL..95w5003L}
N.~F. {Loureiro}, S.~C. {Cowley}, W.~D. {Dorland}, M.~G. {Haines}, A.~A. {Schekochihin}, {X-Point Collapse and Saturation in the Nonlinear Tearing Mode Reconnection}, \prl 95~(23) (2005) 235003.
\newblock \href {http://arxiv.org/abs/physics/0507206} {\path{arXiv:physics/0507206}}, \href {https://doi.org/10.1103/PhysRevLett.95.235003} {\path{doi:10.1103/PhysRevLett.95.235003}}.

\bibitem{2007PhPl...14j0703L}
N.~F. {Loureiro}, A.~A. {Schekochihin}, S.~C. {Cowley}, {Instability of current sheets and formation of plasmoid chains}, Physics of Plasmas 14~(10) (2007) 100703--100703.
\newblock \href {http://arxiv.org/abs/astro-ph/0703631} {\path{arXiv:astro-ph/0703631}}, \href {https://doi.org/10.1063/1.2783986} {\path{doi:10.1063/1.2783986}}.

\bibitem{1978mfge.book.....M}
H.~K. {Moffatt}, {Magnetic field generation in electrically conducting fluids}, 1978.

\bibitem{1975JFM....68..769F}
U.~{Frisch}, A.~{Pouquet}, J.~{Leorat}, A.~{Mazure}, {Possibility of an inverse cascade of magnetic helicity in magnetohydrodynamic turbulence}, Journal of Fluid Mechanics 68 (1975) 769--778.
\newblock \href {https://doi.org/10.1017/S002211207500122X} {\path{doi:10.1017/S002211207500122X}}.

\bibitem{2022JPlPh..88f9002Z}
H.~{Zhou}, R.~{Sharma}, A.~{Brandenburg}, {Scaling of the Hosking integral in decaying magnetically dominated turbulence}, Journal of Plasma Physics 88~(6) (2022) 905880602.
\newblock \href {http://arxiv.org/abs/2206.07513} {\path{arXiv:2206.07513}}, \href {https://doi.org/10.1017/S002237782200109X} {\path{doi:10.1017/S002237782200109X}}.

\bibitem{2023Atmos..14..932B}
A.~{Brandenburg}, G.~{Larsson}, {Turbulence with Magnetic Helicity That Is Absent on Average}, Atmosphere 14~(6) (2023) 932.
\newblock \href {http://arxiv.org/abs/2305.08769} {\path{arXiv:2305.08769}}, \href {https://doi.org/10.3390/atmos14060932} {\path{doi:10.3390/atmos14060932}}.

\bibitem{2023JPlPh..89f9006B}
A.~{Brandenburg}, R.~{Sharma}, T.~{Vachaspati}, {Inverse cascading for initial magnetohydrodynamic turbulence spectra between Saffman and Batchelor}, Journal of Plasma Physics 89~(6) (2023) 905890606.
\newblock \href {http://arxiv.org/abs/2307.04602} {\path{arXiv:2307.04602}}, \href {https://doi.org/10.1017/S0022377823001253} {\path{doi:10.1017/S0022377823001253}}.

\bibitem{2024arXiv240611798B}
A.~{Brandenburg}, A.~{Banerjee}, {Turbulent magnetic decay controlled by two conserved quantities}, arXiv e-prints (2024) arXiv:2406.11798\href {http://arxiv.org/abs/2406.11798} {\path{arXiv:2406.11798}}, \href {https://doi.org/10.48550/arXiv.2406.11798} {\path{doi:10.48550/arXiv.2406.11798}}.

\bibitem{1991A&A...246...49B}
C.~{Burigana}, L.~{Danese}, G.~{de Zotti}, {Formation and evolution of early distortions of the microwave background spectrum - A numerical study}, \aap 246~(1) (1991) 49--58.

\bibitem{2022PhRvD.106f3527B}
F.~{Bianchini}, G.~{Fabbian}, {CMB spectral distortions revisited: A new take on {\ensuremath{\mu}} distortions and primordial non-Gaussianities from FIRAS data}, \prd 106~(6) (2022) 063527.
\newblock \href {http://arxiv.org/abs/2206.02762} {\path{arXiv:2206.02762}}, \href {https://doi.org/10.1103/PhysRevD.106.063527} {\path{doi:10.1103/PhysRevD.106.063527}}.

\bibitem{1996ApJ...473..576F}
D.~J. {Fixsen}, E.~S. {Cheng}, J.~M. {Gales}, J.~C. {Mather}, R.~A. {Shafer}, E.~L. {Wright}, {The Cosmic Microwave Background Spectrum from the Full COBE FIRAS Data Set}, \apj 473 (1996) 576.
\newblock \href {http://arxiv.org/abs/astro-ph/9605054} {\path{arXiv:astro-ph/9605054}}, \href {https://doi.org/10.1086/178173} {\path{doi:10.1086/178173}}.

\bibitem{2013A&ARv..21...62D}
R.~{Durrer}, A.~{Neronov}, {Cosmological magnetic fields: their generation, evolution and observation}, \aapr 21 (2013) 62.
\newblock \href {http://arxiv.org/abs/1303.7121} {\path{arXiv:1303.7121}}, \href {https://doi.org/10.1007/s00159-013-0062-7} {\path{doi:10.1007/s00159-013-0062-7}}.

\bibitem{Brandenburg:2024tyi}
A.~Brandenburg, A.~Neronov, F.~Vazza, {Resistively controlled primordial magnetic turbulence decay}, Astron. Astrophys. 687 (2024) A186.
\newblock \href {http://arxiv.org/abs/2401.08569} {\path{arXiv:2401.08569}}, \href {https://doi.org/10.1051/0004-6361/202449267} {\path{doi:10.1051/0004-6361/202449267}}.

\bibitem{2017JCAP...12..011J}
D.~{Jim{\'e}nez}, K.~{Kamada}, K.~{Schmitz}, X.-J. {Xu}, {Baryon asymmetry and gravitational waves from pseudoscalar inflation}, \jcap 2017~(12) (2017) 011.
\newblock \href {http://arxiv.org/abs/1707.07943} {\path{arXiv:1707.07943}}, \href {https://doi.org/10.1088/1475-7516/2017/12/011} {\path{doi:10.1088/1475-7516/2017/12/011}}.

\bibitem{1956pfig.book.....S}
L.~{Spitzer}, {Physics of Fully Ionized Gases}, 1956.

\bibitem{1950PhRv...80..230C}
R.~S. {Cohen}, L.~{Spitzer}, P.~M. {Routly}, {The Electrical Conductivity of an Ionized Gas}, Physical Review 80~(2) (1950) 230--238.
\newblock \href {https://doi.org/10.1103/PhysRev.80.230} {\path{doi:10.1103/PhysRev.80.230}}.

\bibitem{2017ipp..book.....G}
D.~A. {Gurnett}, A.~{Bhattacharjee}, {Introduction to Plasma Physics}, 2017.

\bibitem{1953PhRv...89..977S}
L.~{Spitzer}, R.~{H{\"a}rm}, {Transport Phenomena in a Completely Ionized Gas}, Physical Review 89~(5) (1953) 977--981.
\newblock \href {https://doi.org/10.1103/PhysRev.89.977} {\path{doi:10.1103/PhysRev.89.977}}.

\bibitem{2003phpl.book.....B}
T.~J.~M. {Boyd}, J.~J. {Sanderson}, {The Physics of Plasmas}, 2003.
\newblock \href {https://doi.org/10.1017/CBO9780511755750} {\path{doi:10.1017/CBO9780511755750}}.

\bibitem{2022JPlPh..88e1501S}
A.~A. {Schekochihin}, {MHD turbulence: a biased review}, Journal of Plasma Physics 88~(5) (2022) 155880501.
\newblock \href {http://arxiv.org/abs/2010.00699} {\path{arXiv:2010.00699}}, \href {https://doi.org/10.1017/S0022377822000721} {\path{doi:10.1017/S0022377822000721}}.

\end{thebibliography}
\end{document}